# A multi-band, multi-level multi-electron model for efficient FDTD simulations of electromagnetic interactions with semiconductor quantum wells

Koustuban Ravi[1,3]*, Qian Wang[1] and Seng-Tiong Ho[2]

*1. Data Storage Institute, Agency of Science Technology and Research, National University of Singapore, 117608 Singapore*

*2. Northwestern University, Evanston, IL,USA*

*3. (This work was completed in Data Storage Institute. The author is currently affiliated with the Massachusetts Institute of Technology, Cambridge, MA,USA)*

*Research Laboratory of Electronics, 50 Vassar Street, Cambridge, Massachusetts, 02139, USA.

email :koust@mit.edu

# A multi-band, multi-level multi-electron model for efficient FDTD simulations of electromagnetic interactions with semiconductor quantum wells

We report a new computational model for simulations of electromagnetic interactions with semiconductor quantum well(s) (SQW) in complex electromagnetic geometries using the finite difference time domain (FDTD) method. The presented model is based on an approach of spanning a large number of electron transverse momentum states in each SQW sub-band (multi-band) with a small number of discrete multi-electron states (multi-level, multi-electron). This enables accurate and efficient two dimensional (2-D) and 3-D simulations of nanophotonic devices with SQW active media. The model includes the following features: (1) Optically induced interband transitions between various SQW conduction and heavy-hole or light-hole sub-bands are considered. (2) Novel intra sub-band and inter sub-band transition terms are derived to thermalize the electron and hole occupational distributions to the correct Fermi-Dirac distributions. (3) The terms in (2) result in an explicit update scheme which circumvents numerically cumbersome iterative procedures. This significantly augments computational efficiency. (4) Explicit update terms to account for carrier leakage to unconfined states are derived which thermalize the bulk and SQW populations to a common quasi-equilibrium Fermi-Dirac distribution. (5) Auger recombination and intervalence band absorption are included. The model is validated by comparisons to analytic band filling calculations, simulations of SQW optical gain spectra and photonic crystal lasers.

Keywords: Finite difference time domain method, Nanophotonic Devices

## 1. Introduction

The finite difference time domain (FDTD) method [1] is a numerical approach for the solution of Maxwell's equations in complex electromagnetic structures. It has emerged as a powerful and general method for electromagnetic simulations of nanophotonic devices of complex electromagnetic geometry. Recently, there is an interest in applying FDTD to simulate active photonic devices (where there is amplification, absorption or emission of light). This is achieved by combining the standard FDTD method with suitable active medium models to govern the light-matter interaction physics [2]-[12] (active media include atomic, molecular or semiconductor materials). Of various active media, semiconductor quantum well(s) (SQW) are among the most commonly used in optoelectronic and nanophotonic devices (henceforth referred to collectively as nanophotonic devices).

Hence, if a physically accurate active medium model for SQW is combined with FDTD, it could be extremely powerful for understanding and developing insights into a wide range of nanophotonic devices of complex electromagnetic geometry.

However, since FDTD can be computationally demanding owing to the use of sub-wavelength spatial resolution, the active medium model must be formulated to be computationally efficient in addition to being physically accurate. Otherwise, simulations would be forbiddingly time consuming. In this paper, we set about the task of formulating an active medium model to govern light–matter interaction with SQW accurately, as well as efficiently enough, to enable FDTD simulations of nanophotonic devices of complex two dimensional (2-D)/3-D electromagnetic geometry.

Existing active medium FDTD implementations include those based on two-level [2] or four–level [3]-[6] atomic systems. However, in order to account for the correct optical gain (henceforth called gain) dispersions and band-filling effects, the continuous band structure of semiconductors has to be accounted for. Other active medium FDTD models use multiple dipoles to account for the correct gain dispersion but use conventional quasi-equilibrium carrier rate equations [7]-[11] to govern the carrier dynamics. Carrier rate equation based approaches [7]-[11] are valid when the distributions do not deviate significantly from Fermi-Dirac functions. Furthermore, they contain phenomenological parameters which are highly dependent on operating conditions or device geometry.

For greater generality, one would have to self-consistently track the dynamics of the complete carrier occupational distributions along with the dynamics of the dipoles and electromagnetic fields. A case in point for the need for such a self-consistent calculation is when optical fields with multiple spectral components of high intensity simultaneously interact with the active medium. In such a case, each spectral component modifies the carrier occupational distribution which in turn affects the interaction of every other spectral component.

Accurate and widely applicable non-equilibrium models based on Maxwell-Bloch equations [12]-[15] have been developed which provide such detailed treatment of SQW carrier dynamics and band structure. Some of these models have even been combined with 3-D FDTD simulations as in [12]. However, such methods usually involve iterative procedures to update chemical potentials at each spatial location and time instant, which can be computationally quite intensive.

In this paper we propose a new model to govern optical interactions with SQW both accurately and efficiently to enable 2-D/3-D simulations of Nanophotonic devices. The presented model straddles a middle-ground between the phenomenological carrier rate equation approaches [7]-[11] and the *ab-initio* Maxwell-Bloch equation approaches [12] in accuracy and computational efficiency. It is based on a Multi- Band, Multi-Level,

Multi-Electron (MB-MLME) system. In the MB-MLME model, we use a series of bands (hence, multi-band) to represent the various SQW sub-bands separately. Each sub-band *l* is then spanned by a few broadened multi-electron states instead of finely resolved transverse momentum states ($k_t$). Since the number of sub-bands are typically few in number (~ 2 conduction/heavy hole/light hole bands), whereas the number of $k_t$ states are much larger in number, large computational savings are achieved which enables FDTD simulations of nanophotonic devices of complex 2-D/3-D electromagnetic geometries to become feasible. The multi-band treatment allows us to consider optically induced interband transitions between various conduction and heavy-hole as well as light-hole sub-bands while obeying the correct selection rules. Furthermore, the difference in light-matter interaction properties for different electromagnetic field components (gain anisotropy) can also be considered.

In our MB-MLME model, we introduce a detailed model to govern the various intra sub-band and inter sub-band transitions. A novel set of transition terms are introduced which thermalize carriers within each sub-band as well between various sub-bands to the correct quasi-equilibrium Fermi-Dirac distributions. Most importantly, the proposed approach circumvents all iterative calculations. Thus an explicit update scheme is developed which augments the computational efficiency of the approach. A similar approach was proposed in [16]-[18] for bulk semiconductors. In [19], a preliminary model which performed a simplistic treatment of the SQW band structure, without considering sub-bands, was presented.

Additionally, we also introduce carrier leakage to various unconfined well, barrier and separate confinement heterostructure (SCH) bulk states. These carrier leakage terms can account for carrier capture and emission processes. They ensure that the well and bulk populations are automatically thermalized to a common quasi-equilibrium Fermi-

Dirac distribution without necessitating any iterative procedures. Finally, Auger recombination and intervalence band absorption are also incorporated phenomenologically for the sake of completeness.

The accuracy of the approach is verified by the excellent agreement between carrier occupational distributions obtained from MB-MLME model simulations to analytic band filling calculations. Furthermore, FDTD simulations of SQW gain spectra show well known characteristics such as the step-like dispersion and agree quantitatively with experimental results [20]. The feasibility of using the MB-MLME model for simulations of devices of complex structural geometry is demonstrated by 2D FDTD simulations of photonic crystal waveguide lasers.

The organization of the paper is as follows. In Section 2, we briefly describe the concept of SQW to keep the paper self contained. In Section 3, we outline the key ideas underlying the MB-MLME model. In Section 4, we present the polarization equations of motion for the various electric dipoles. In Section 5, we present calculations of the dipole number densities. In Section 6, we outline the carrier rate equations for the various sub-bands. Section 7 is a key section where we present the detailed intra sub-band, inter sub-band and barrier leakage terms which automatically thermalize the excited carriers to a common quasi-equilibrium Fermi-Dirac distribution. In Section 8, we show how to couple the active medium model to the Maxwell's equations. In Section 9, we present computationally efficient explicit FDTD update equations for the MB-MLME model. In section 10.1, we verify the MB-MLME model against analytic band filling calculations. In section 10.2, we present simulations of SQW gain spectra and contrast it to bulk semiconductor gain spectra. In section 10.3, we apply our MB-MLME FDTD approach to simulate 2D photonic crystal waveguide lasers. In Section

10.4, we present results indicating computational efficiency. We conclude in section 11. Appendices are provided to keep the paper self contained.

## 2. Description of SQW band structures

In this section, we briefly describe the band structure of SQW for completeness. SQW are formed by a semiconductor of band-gap $E_G^{qw}$ with thickness $L^{qw}$ between semiconductor layers with larger band-gap $E_G^B$. For all parameters, superscripts 'qw' and 'B' are used to denote properties of the quantum well and barrier regions respectively. These superscripts are not to be confused as variables raised to some power. Within the barrier and well regions, the electrons experience a periodic potential due to the periodic semiconductor crystal structure. The effect of these periodic potentials can be described by effective mass parameters. In this effective mass approximation, the electrons see a square potential well with depth $\Delta E_{C-offset}$ as shown in Fig.1. In all parameters henceforth, the subscript 'C' refers to the conduction band. Similarly, subscripts 'HH' and 'LH' refer to the heavy-hole and light-hole respectively. In case, the degeneracy of holes is ignored, a subscript 'V' is used to denote the valence band.

Holes see a finite potential well of depth $\Delta E_{V-offset} = \left(E_G^B - E_G^{qw}\right) - \Delta E_{C-offset}$ with corresponding effective masses $m_{HH\perp}^{qw}$ and $m_{HH}^B$ for heavy-holes (HH) and $m_{LH\perp}^{qw}$ and $m_{LH}^B$ for light-holes (LH). The symbol '$\perp$' in the subscripts refer to properties out of ($y$ direction) of the quantum well respectively. The effect of the SQW structure is similar to the 'particle in a box' problem [21] which results in the quantization of the possible energy states an electron or hole can occupy. This is depicted in Fig.1, where the horizontal blue lines indicate the first quantized conduction ($E_{C(l)}^{qw-0}$), heavy-hole ($E_{HH(l)}^{qw-0}$) and light-hole ($E_{LH(l)}^{qw-0}$) quantized states. Here, the subscript $LH(l)$ corresponds to the $l^{th}$ SQW light-hole sub-band. The super-script 'qw-0' corresponds to the quantum well sub-band edge. The light-hole states are represented by dashed lines to distinguish them from the heavy-hole states. Similarly, the red lines indicate the corresponding second quantized states. The energy separation between the first quantized conduction band state and first quantized HH state is defined as $E_{C(1)-HH(1)}$ and so on. The number of quantized states will depend on the well width and depth.

However, in directions parallel to the well layer ($x$ and $z$ in Fig.1), electrons and holes behave like 'free particles'. As a result, the total energy for electrons in the $l^{th}$ quantized state $E_{C(l)}^{qw}$ can be written in the form $E_{C(l)}^{qw}(k_t) = E_{C(l)}^{qw-0} + (2m_{C\|(l)}^{qw})^{-1}\hbar^2 k_t^2$, where $E_{C(l)}^{qw-0}$ is the energy of the $l^{th}$ conduction band quantized energy state and is analogous to the potential energy. Here, $m_{C(l)\|}^{qw}$ is the in-plane effective electron mass for an electron in the $l^{th}$ quantized state. The symbol '$\|$' in the subscripts above accounts for the net effect of the semiconductor crystal in the in-plane directions ($x$ and $z$ directions). Similar arguments hold for the valence bands with $m_{HH\|(l)}^{qw}$ and $m_{LH\|(l)}^{qw}$ being in-plane effective masses of the $l^{th}$ heavy-hole and light-hole sub-bands. Thus, each quantized state is associated with different parabolic energy-momentum dispersions (In Fig.2 the $l^{th}$ sub-band is represented by the same color as the $l^{th}$ quantized state in Fig.1). Overall, the band structure of SQW's are a series of sub-bands as shown in Fig.2.

Note that the parabolic description is presented here to delineate the main concept of SQW with simplicity. In reality, there may be multiple SQW as well as non-parabolicity and asymmetry. However, the MB-MLME model formulations will be valid for many band structures. For instance, the different potential energy functions of type I and type II heterojunctions are reflected in the values of the quantized energies $E_{C(l)}^{qw-0}$ and effective mass parameters.

## 3. MB-MLME approach to modelling quantum wells

### 3.1 Multi-Band, Multi-Level, Multi-electron scheme of energy levels to modelling optical interaction with SQW

Typically, in SQW, the number of conduction, heavy-hole or light-hole sub-bands $l$ are few (~2-3 each) but the number of $k_t$ states are infinitely large. Each $k_t$ state in the $l^{th}$ sub-band is associated with an electric dipole moment and carrier occupancy parameter. Thus, in general the dynamics of dipoles and carrier populations have to be tracked over a very large number of $k_t$ states. Since in FDTD simulations, there can be millions of spatial locations filled with SQW active medium, such computation of carrier and dipole dynamics over a large range of finely resolved $k_t$ states is too forbiddingly time consuming for the simulation of complex 2-D/3-D Nanophotonic devices.

In the MB-MLME approach, we introduce the concept of representing a group of $k_t$ states within each sub-band by a single, broadened multi-electron state. The rationale for this is based on the notion that dipoles in semiconductor media undergo ultrafast dephasing processes which leads to large broadening. Thus a broad bandwidth of energy within each sub-band can be spanned by just few multi-electron states instead of finely resolved momentum states. This can significantly alleviate the computational burden of accurately simulating light-matter interaction with SQW. A similar concept was developed in [16] only for bulk semiconductors. Since, in the MB-MLME model each sub-band is treated separately, in addition to having multiple multi-electron states, we will also have multiple sub-bands. Hence, we call this approach the Multi-Band, Multi-Level, Multi-Electron (MB-MLME) approach.

The scheme of the MB-MLME model is depicted in Fig.3. A range of electron states in the energy bracket $[E_{C(l,i)}^{qw-}, E_{C(l,i)}^{qw+}]$ within the $l^{th}$ conduction sub-band (indicated by the shaded regions in the conduction bands of Fig.3) are represented by a single, broadened multi-electron state $|l,i\rangle_C$ (black, dashed line). Thus a series of states $|l,1\rangle_C..|l,i\rangle_C,|l,i+1\rangle_C..$ are used to represent the entire range of electron states in the $l^{th}$ conduction sub-band. Physically, the size of the energy bracket $[E_{C(l,i)}^{qw-}, E_{C(l,i)}^{qw+}]$ spanned by each broadened multi-electron state must be on the order of the dipole broadening caused by dephasing due to carrier-carrier scattering. The energy broadening due to scattering processes is approximately given by $\Delta E_t = 2\hbar \tau_{scat}^{-1}$ [20]. In terms of wavelength, this is given by a wavelength interval $\Delta \lambda = 2c^{-1}\lambda_n^2 \tau_{scat}^{-1}$, where $\lambda_n$ is the wavelength of light in the material. For III-V materials, with band-gaps ~1μm and refractive indices ~ 3 this corresponds to a $\Delta \lambda$ of about 20 nm. If the wavelength interval is much larger than the above value, the gain/absorption spectrum shows significant corrugations as shown in [16]. For energy spacing much smaller than the above value, the number of levels increases and computational efficiency is lost. In addition for such fine spacing, the dynamics of carriers are slowed down due to a larger number of transitions. Similarly, a series of multi-electron states $|l,i\rangle_{HH}$ (green, dash) and $|l,i\rangle_{LH}$ (blue, dash) represent all electron states in the energy brackets $[E_{HH(l,i)}^{qw-}, E_{HH(l,i)}^{qw+}]$ and $[E_{LH(l,i)}^{qw-}, E_{LH(l,i)}^{qw+}]$ respectively in the $l^{th}$ heavy-hole and light-hole sub-bands. Here the superscripts '-' and '+' represent, the lower and upper limit of the energy bracket respectively. The reader is referred to Appendix C for a summary of all the parameters. The ensemble of the various $l$ sub-bands then constitutes the overall system as shown in Fig.3.

In Fig.3, the various $l^{th}$ conduction, heavy-hole and light-hole sub-bands are grouped together followed by $l+1^{th}$ and so on. The complete ensemble of all the $l$ sub-bands is thus used to govern the light-matter interaction with SQW. The grouping according to the same $l$ value allows us to satisfy the correct selection rules for transitions in SQW. Selection rules for optically induced electronic transitions occur in accordance with the conservation of momentum and energy. Thus, the change in the transverse momentum of an electron making a transition from a light-hole/heavy-hole band to the conduction band is $\Delta k_t = 0$. The $\Delta k_t = 0$ rule is satisfied by appropriately setting the limits of energy brackets $[E_{HH(l,i)}^{qw-}, E_{HH(l,i)}^{qw+}]$ and $[E_{LH(l,i)}^{qw-}, E_{LH(l,i)}^{qw+}]$ in accordance with $[E_{C(l,i)}^{qw-}, E_{C(l,i)}^{qw+}]$ (See Appendix A).

Additionally, transitions occur only between a conduction and hole sub-band with the same sub-band index $l$, i.e $\Delta l = 0$ [22]. This is because the dipole matrix elements for such transitions is zero due to a vanishing overlap integral of the wave-functions [22]. Thus, the dipole-moment $\bar{\mu}_{C-HH(l,i)}^{qw}$ associated with the pair of states- $|l,i\rangle_C, |l,i\rangle_{HH}$ covers all transitions between the $l^{th}$ conduction band and heavy-hole bands in the transition energy bracket $[E_{C(l,i)}^{qw-}, E_{C(l,i)}^{qw+}]$, centered at $E_{C-HH(l,i)}^{qw}$ as shown in Fig.3. The transition energy $E_{C-HH(l,i)}^{qw} = E_{C(l,i)}^{qw} - E_{HH(l,i)}^{qw}$, where $E_{C(l,i)}^{qw}$ is the energy of the state $|l,i\rangle_C$ (w.r.t vacuum) and $E_{HH(l,i)}^{qw}$ is the energy of the state $|l,i\rangle_{HH}$ as shown in Fig.2. Similarly, the dipole-moment $\bar{\mu}_{C-LH(l,i)}^{qw}$ linking the pair of states- $|l,i\rangle_C, |l,i\rangle_{LH}$ encompasses all transitions between the $l^{th}$ conduction and light-hole sub-bands in the transition energy bracket $[E_{C-LH(l,i)}^{qw-}, E_{C-LH(l,i)}^{qw+}]$ centred at $E_{C-LH(l,i)}^{qw}$. By treating each sub-band separately, we are thus able to model the optical interaction with SQW's over a broad bandwidth while automatically accounting for the $\Delta k_t = 0, \Delta l = 0$ selection rules.

### 3.2 Developing computationally efficient transition schemes for Thermalizing carrier occupational distributions in SQW

Typically, the process of determining the correct Fermi-Dirac distributions involves cumbersome iterative calculations of chemical potentials and carrier temperatures based on carrier number density and carrier kinetic energy density conservation. However, since 2-D/3-D FDTD simulations can have millions of spatial pixels, such iterative calculations at each grid point and each time instant would be computationally expensive. We address this problem by devising a novel scheme of carrier transitions for SQW using the concept of energy up transitions (EUT) and energy down transitions (EDT) developed in [16]-[17]. However the works in [16]-[17] only developed a scheme of transitions to thermalize carriers in bulk semiconductors. In [17], it was shown that when Pauli's exclusion is considered and when intraband EUT transition times $\tau_u$ and intraband EDT transition times $\tau_d$ between two levels separated by energy $\Delta E$ are related such that $\tau_u \propto \tau_d \exp\{\Delta E / k_B T_P\}$ (See Appendix.B for derivation), carrier occupational distributions are automatically thermalized to the correct Fermi-Dirac distributions at the corresponding plasma (electron or hole) temperature $T_p$. This approach circumvents the need to calculate chemical potentials altogether and provides an explicit update scheme for the carrier rate equations which results in high computational efficiency for FDTD simulations. The EUT and EDT emulate carrier-carrier scattering processes [17], when driven by the plasma temperature $T_p$ (along with a rate equation for $T_p$). If we fix $T_p$ to the lattice temperature $T_L$, they resemble carrier-phonon scattering processes. In SQW's, the problem of thermalizing carriers is more complex as carriers in all the sub-bands have to be thermalized to a single Fermi-Dirac distribution at a common temperature. This would require carriers to not only be thermalized within each sub-band but also between various sub-bands. We thus introduce a scheme of intra sub-band transitions (Fig.4a) as well as inter sub-band transition terms (Fig.4b) in order to do so.

*3.2.1. Intra sub-band and Inter sub-band transitions*

Figure 4a presents the scheme of intra sub-band transitions within the $l^{th}$ conduction sub-band. Electrons in the $i^{th}$ energy level of the $l^{th}$ conduction sub-band can make EDT (blue arrow) to the $(i-1)^{th}$ level with a transition time $\tau_d = \tau_{(l,[i,i-1])C}$. Similarly, an electron in the $(i-1)^{th}$ level can make EUT (red arrow) to the $i^{th}$ level with transition time $\tau_u = \tau_{(l,[i-1,i])C}$. Since, each sub-band thermalizes to a different carrier temperature at first, the ratio of EUT and EDT should be set as $\tau_{(l,[i-1,i])C} \propto \tau_{(l,[i,i-1])C} \exp\{(E^{qw}_{C(l,i)} - E^{qw}_{C(l,i)})/k_B T^{qw}_{e(l)}\}$, where $T^{qw}_{e(l)}$ electron temperature of the $l^{th}$ SQW sub-band and $E^{qw}_{C(l,i)}$ is the energy (w.r.t. vaccum) of the $i^{th}$ energy level in the $l^{th}$ conduction sub-band. Note, that in our current formulation, we include transitions only between adjacent levels $i, i-1$ within the $l^{th}$ sub-band. However, we may also consider transitions between any two levels $i, j$ without changing the approach.

In order to thermalize the various sub-bands to a common Fermi-Dirac distribution, we also introduce inter sub-band transitions as depicted in Fig.4b. Carrier-phonon scattering is typically responsible for inter sub-band transitions between the bottom of any two sub-bands $l,m$ [23]. Thus, we set EDT and EUT between the first energy levels in the $l^{th}$ and $m^{th}$ ($l>m$) conduction sub-bands according to $\tau_{([m,l],1)C} \propto \tau_{([l,m],1)C} \exp\{(E^{qw}_{C(l,1)} - E^{qw}_{C(m,1)})/k_B T_L\}$ as shown in Fig.4b.

Thus, while each sub-band thermalizes independently to a different carrier temperature $T^{qw}_{e(l)}$ through the intra-band transition scheme in Fig.4a, they also thermalize to a common lattice temperature $T_L$ through the inter sub-band transitions in Fig.4b to a common lattice temperature $T_L$. Thus, in quasi-equilibrium all carriers will be thermalized to a single Fermi-Dirac distribution at the lattice temperature and a common chemical potential. We verify in Section 10.1, that this is indeed the case. For many applications where critical time-scales exceed picoseconds, this thermalization to multiple temperatures may not be important. This includes continuous wave lasing, picosecond pulse amplification for a variety of electromagnetic structures. Therefore, for simplicity in this paper, we do not track the dynamics of the electron temperatures $T^{qw}_{e(l)}(\mathbf{r},t)$ and simply set all $T^{qw}_{e(l)}(\mathbf{r},t) = T_L$. However, for systems where femtosecond timescales become important such as femtosecond pulse amplification, one would have to consider these non-equilibrium phenomena. A similar approach is used to thermalize the occupational distribution of heavy-holes/ light-holes.

*3.2.2. Modeling carrier leakage to bulk states*

In SQW systems, in addition to thermalizing carriers within various SQW states, it is necessary to account for the leakage of carriers to the bulk states which include unconfined well, barrier and even separate confinement heterostructure (SCH) states. This is because these bulk states can have important implications on device performance [24]. In the MB-MLME approach, we use the same concept of EUT and EDT between bulk states and any $l^{th}$ conduction sub-band as shown in Fig.5 to thermalize the carrier populations in both SQW and bulk states to a common quasi-equilibrium Fermi-Dirac distribution. In, Fig.5 EDT (blue) with transition time $\tau_d = \tau_{B \to C(l)}$ occurs between the bottom of the bulk bands and the bottom of the $l^{th}$ conduction sub-band of the SQW. The subscript 'B' here is used to denote bulk states. The red arrows represent EUT with transition time $\tau_u = \tau_{C(l) \to B}$. Similar to the inter sub-band and intra sub-band transitions, by setting $\tau_{C(l) \to B} \propto \tau_{B \to C(l)} \exp\left\{(E_{C(1)}^{B} - E_{C(l,1)}^{qw})/k_B T_L\right\}$ and including Pauli's exclusion principle, we can thermalize the SQW and bulk populations to a common Fermi-Dirac distribution. In essence, the concept of EDT and EUT applied to carrier leakage is equivalent to the process of carrier capture and escape as indicated in Fig.5. In fact, the ratio of carrier capture and escape times have previously contained the exponential factor based on a thermionic emission model [25]. However, such previous work did not include Pauli's exclusion and therefore did not yield the correct Fermi-Dirac distributions. This approach is an approximation since the non-uniform distribution of carriers between wells is not considered via spatial transport in the multiple well-barrier system but serves as an approximate approach to determine the distribution of carriers in a multiple well system.

### *3.3. An effective medium approach for FDTD simulations of optical interaction with SQW*

The stability of FDTD [34] requires the ratio of spatial and temporal resolutions to be much smaller than the speed of light for sufficient accuracy, i.e. $\Delta x/\Delta t \ll c$. Since, the full electromagnetic field is updated, a $\Delta t$ much smaller than the time period of the electromagnetic wave is required. Correspondingly, the spatial resolution must be much smaller than the optical wavelength. Therefore, a typical grid size used is $\sim \lambda_n (20)^{-1}$ [16], where $\lambda_n$ is the wavelength in the material which is about ~10nm at optical frequencies. Since this is typically larger than the SQW width itself (which is usually <10 nm), we use an effective medium approach to model the SQW and barrier regions using FDTD. In this case, each grid in the active region is uniformly filled with an effective band structure accounting for both the SQW's as well as barriers as shown in Fig.6.

### *3.4. MB-MLME model vs Actual Physical situation*

While an attempt to cover the essential aspects of light interaction with SQW has been made in the formulation of the MB-MLME model, there still exist certain approximations in relation to actual physical situations. Firstly, in the presented formulation of the MB-MLME model, we assume that transitions only occur between sub-bands of the same *l* value.

In reality, transitions can occur between any two sub-bands albeit, with significantly reduced strengths when $\Delta l \neq 0$ [20].

While it is possible to include additional dipoles to govern these 'forbidden' transitions, they are omitted here for the sake of greater computational efficiency.

Secondly, in the various thermalization processes, we assume that the various intra sub-band, inter sub-band and carrier leakage EDT times are constant. In reality, these relaxation times are complex functions of carrier density, temperature as well as the functional form of the carrier distribution. Even the transition schemes in Sections 3.2 are significantly simplified. In reality one would have to consider scattering events between any two states and self-consistently calculate the scattering rates. Such, highly accurate models would then require a self-consistent calculation of these terms as in Monte –Carlo simulations [26]. However, the complex calculations involved would make it forbiddingly time consuming for full scale FDTD simulation of devices.

An alternative solution would be to pre-compute these transition times for various conditions and develop a look up table for dynamic update during the simulations. Such approaches have been used in [15]. Additionally, many-body effects such as band gap renormalization and coulomb enhancements of dipole matrix elements affect SQW gain spectra [27]-[28] and not currently included. Finally, excitonic effects which are important for low temperature device simulations are currently not included. Excitonic effects maybe include by introducing energy levels and modified matrix elements as in [36]. The coulomb enhancement and bandgap renormalization effects may be included using expressions such as Eqs. (25)-(30) from [40].

## 4. Polarization equations of motion including gain anisotropy

### 4.1 Polarization Equations of Motion for various SQW and bulk states

In Section 3.1, we described the general scheme of the MB-MLME mode of representing several transverse momentum ($k_t$) electron states in the $l^{th}$ conduction and heavy-hole sub-bands by a finite number of multi-electron broadened states - $|l,i\rangle_C$ and $|l,i\rangle_{HH}$ respectively. An electric dipole-moment $\vec{\mu}_{C-HH(l,i)}^{qw}$ centered at the transition energy $E^{qw}_{C-HH(l,i)}$ then governs all optical transitions in the transition energy bracket $[E^{qw-}_{C-HH(l,i)}, E^{qw+}_{C-HH(l,i)}]$ (See Fig.3) spanned by the pair of broadened multi-electron states $|l,i\rangle_C$ and $|l,i\rangle_{HH}$. If the number of conduction sub-bands and heavy-hole sub-bands are not equal, dipole moments are setup only between those conduction and heavy-hole sub-bands for which $\Delta l = 0$ in our current implementation. This is a very good approximation [20] although the approach can be extended to include transitions between sub-bands for which $\Delta l \neq 0$. Thus, $\vec{\mu}^{qw}_{C-HH(l,i)}$ denotes the dipole-moment corresponding to the $i^{th}$ pair of broadened states in the $l^{th}$ SQW conduction and heavy-hole sub-bands. Since, the dipole moment $\vec{\mu}^{qw}_{C-HH(l,i)}$ acts as an electromagnetic field source, it alters the total electric field $\vec{E}(\mathbf{r},t)$ at each position vector $\mathbf{r}$ and time instant $t$ via an electromagnetic polarization density vector $\vec{P}^{qw}_{C-HH(l,i)}(\mathbf{r},t)$ in the magnetic curl equation (See. Eq. (23), Section 8).

The microscopic dipole moment, $\vec{\mu}^{qw}_{C-HH(l,i)}$ spans all optical interactions in the energy bracket $[E^{qw-}_{C-HH(l,i)}, E^{qw+}_{C-HH(l,i)}]$. Therefore, the polarization density $\vec{P}^{qw}_{C-HH(l,i)}(\mathbf{r},t)$ at each $\mathbf{r}$ is the microscopic dipole moment $\vec{\mu}^{qw}_{C-HH(l,i)}$ multiplied by the total number of dipoles in the transition energy bracket $[E^{qw-}_{C-HH(l,i)}, E^{qw+}_{C-HH(l,i)}]$, in a small volume $\delta V$ centered at $\mathbf{r}$. Since, the entire barrier and SQW regions will be represented by an effective medium (See 3.3) uniformly filling the active region at each $\mathbf{r}$, we can represent $\delta V = L_{ac} \delta A$, where $L_{ac}$ is the total active region thickness (of barriers and wells) and $\delta A$ is a small area in the plane of the wells. Thus, $\vec{P}^{qw}_{C-HH(l,i)} = L_{ac}^{-1} N^{qw}_{dip(l,i)} \vec{\mu}^{qw}_{C-HH(l,i)}$, where $N^{qw}_{dip(l,i)}$ is known as the SQW dipole number density (See Section 5 for calculation). $N^{qw}_{dip(l,i)}$ is the total number of dipoles per unit area in the energy bracket $[E^{qw-}_{C-HH(l,i)}, E^{qw+}_{C-HH(l,i)}]$ spanned by the dipole.

In FDTD, continuous space is discretized into several grids of volume $\delta V$ centered at discretized spatial co-ordinates $\mathbf{r_j}$ and continuous time is discretized into time instants $t_n$. However, we will first present all formulations in continuous variables. The spatial-temporal evolution of $\vec{P}^{qw}_{C-HH(l,i)}(\mathbf{r},t)$ is similar to the polarization equations of motion derived in [16] and is presented in Eq.(1). However, since SQW's display different light-matter interaction behavior for different electric field components (gain anisotropy), we modify the polarization equations of motion to be different for components parallel (i.e. $P^{qw}_{x[C-HH(l,i)]}(\mathbf{r},t), P^{qw}_{z[C-HH(l,i)]}(\mathbf{r},t)$) and perpendicular to the plane of quantum wells (i.e. $P^{qw}_{y[C-HH(l,i)]}(\mathbf{r},t)$) as shown in Eqs. (1a) and (1b) respectively. This is in contrast to bulk materials where the light-matter interaction is isotropic [16]. This anisotropy is incorporated via different dipole-matrix elements $|\mu^{qw}_{C-HH(l,i)\|}|^2$ and $|\mu^{qw}_{C-HH(l,i)\perp}|^2$ respectively. The '$\|$' and '$\perp$' subscripts have the meanings defined earlier. The matrix elements are indicative of the strength of the dipole oscillations. For a quick reference of all variables, the reader is referred to Appendix.C. The various components of the polarization density vector are updated at the same spatial locations as the corresponding electric field components in the standard Yee algorithm for FDTD simulations. The entire active region is uniformly filled with an effective quantum well medium as described in Section 3.3. This approach allows for the use of spatial grid-sizes larger than the quantum well dimensions.

$$\frac{d^2 P^{qw}_{x[C-HH(l,i)]}(\mathbf{r},t)}{dt^2} + \gamma^{qw}_{C-HH(l,i)} \frac{dP^{qw}_{x[C-HH(l,i)]}(\mathbf{r},t)}{dt} + \left[ \left(\omega^{qw}_{C-HH(l,i)}\right)^2 + \frac{4\left(\omega^{qw}_{C-HH(l,i)}\right)^2 \left|\mu^{qw}_{C-HH(l,i)\|}\right|^2}{\hbar^2} A_x^2(\mathbf{r},t) \right] P^{qw}_{x[C-HH(l,i)]}(\mathbf{r},t)$$

$$= \frac{2\left(\omega^{qw}_{C-HH(l,i)}\right)^2}{\hbar} \left|\mu^{qw}_{C-HH(l,i)\|}\right|^2 \frac{1}{L_{ac}} \left[ N^{qw}_{HH(l,i)}(\mathbf{r},t) - N^{qw}_{C(l,i)}(\mathbf{r},t) \right] E_x(\mathbf{r},t)$$

(1a)

$$\frac{d^2 P^{qw}_{y[C-HH(l,i)]}(\mathbf{r},t)}{dt^2} + \gamma^{qw}_{C-HH(l,i)} \frac{dP^{qw}_{y[C-HH(l,i)]}(\mathbf{r},t)}{dt} + \left[\left(\omega^{qw}_{C-HH(l,i)}\right)^2 + \frac{4\left(\omega^{qw}_{C-HH(l,i)}\right)^2 \left|\mu^{qw}_{C-HH(l,i)\perp}\right|^2}{\hbar^2} A_y^2(\mathbf{r},t)\right] P^{qw}_{y[C-HH(l,i)]}(\mathbf{r},t)$$

$$= \frac{2\left(\omega^{qw}_{C-HH(l,i)}\right)^2}{\hbar} \left|\mu^{qw}_{C-HH(l,i)\perp}\right|^2 \frac{1}{L_{ac}} \left[N^{qw}_{HH(l,i)}(\mathbf{r},t) - N^{qw}_{C(l,i)}(\mathbf{r},t)\right] E_y(\mathbf{r},t)$$

(1b)

Equation (1) resembles a harmonic oscillator equation with a resonant angular frequency $\omega^{qw}_{C-HH(l,i)} = \hbar^{-1} E^{qw}_{C-HH(l,i)}$. Thus, an electromagnetic field with angular frequency $\omega^{qw}_{C-HH(l,i)}$ will drive the dipole $\bar{\mu}^{qw}_{C-HH(l,i)}$ the strongest. However, since we use many dipoles in the MB-MLME model, a broadband field with different spectral components would drive dipoles centered at various $\omega^{qw}_{C-HH(l,i)}$. Thus, the MB-MLME approach can be used for simulating broadband interaction with SQW's.

The calculation of $\omega^{qw}_{C-HH(l,i)}$, $E^{qw}_{C-HH(l,i)}$ and other associated energy definitions are dependent on the SQW bandstructure and are briefly described in Appendix A. In Eq. (1), $\gamma^{qw}_{C-HH(l,i)}$ is called the dipole dephasing rate and determines the bandwidth over which an electromagnetic field interacts with a particular dipole. It alludes to the fact that electromagnetic fields detuned by $\hbar \gamma^{qw}_{C-HH(l,i)}/2$ can still excite the dipole with resonant frequency $\omega^{qw}_{C-HH(l,i)}$.

In the RHS of Eqs. (1a) and (1b), we see that the dipoles are driven by the corresponding electric field components as well as the number of electrons per unit area $N^{qw}_{C(l,i)}(\mathbf{r},t)$ and $N^{qw}_{HH(l,i)}(\mathbf{r},t)$ in the corresponding conduction and heavy-hole sub-band levels respectively. Here $N^{qw}_{C(l,i)}(\mathbf{r},t)$ represents the total number of electrons per unit area over the entire energy bandwidth of the broadened state $|l,i\rangle_C$ and varies with both space and time. The temporal dynamics of $N^{qw}_{C(l,i)}(\mathbf{r},t)$ are presented in Section 6 on carrier rate equations. Similarly, $N^{qw}_{HH(l,i)}(\mathbf{r},t)$ is the total number of electrons per unit area in the energy bandwidth spanned by the broadened state $|l,i\rangle_{HH}$ in the heavy-hole band.

The RHS of Eq.(1) contains a factor of $L_{ac}^{-1}$ to average out the polarization density over the entire active medium thickness as mentioned previously. Note that Eq. (1) does not make any assumptions about the band structure [29]. The details of the band structure are accounted for by the various transition energies, resonant frequencies (Appendix.A) and dipole number densities (Section 5). The discussion in this paper is pertinent to direct band-gap semiconductors since they form the active medium of most nanophotonic devices. Silicon, an indirect band-gap material, is mostly used for passive functionality (guiding, routing etc.). However, the MB-MLME model can be used to model indirect band-gap materials by grouping transitions between states which satisfy momentum conservation. Since phonon collisions provide momentum conservation in such systems, energy levels with $\Delta k_t + \Delta k_{phonon} = 0$ rather than $\Delta k_t = 0$ will be clustered together for tracking transitions accurately.

Similar to the polarization density vectors governing transitions between conduction and heavy-hole sub-bands presented in Eq. (1), polarization density vectors $\vec{P}_{C-LH(l,i)}^{qw}(\mathbf{r},t)$ are present for transitions between conduction and light-hole sub-bands. The corresponding polarization equations of motion may be obtained by replacing all the 'HH' subscripts in Eq.(1) to 'LH'.

In the MB-MLME model, in addition to considering SQW states, we also consider optically induced transitions between the conduction and valence bands of bulk unconfined quantum well and barrier states. Once again, to reduce computational time, we span the entire bandwidth of excited carriers in the bulk conduction and valence bands by few broadened multi-electron states $|j\rangle_C$ and $|j\rangle_V$ respectively. The state $|j\rangle_C$ is centered at energy $E^B_{C(j)}$ (w.r.t vacuum level) and $|j\rangle_V$ at energy $E^B_{V(j)}$ (w.r.t vacuum). The pair of states is hence separated by the transition energy $E^B_j = E^B_{C(j)} - E^B_{V(j)}$. The resonant frequency $\omega^B_j = \hbar^{-1} E^B_j$ An electric dipole moment, $\vec{\mu}^B_j(\mathbf{r},t)$ centered at $E^B_j$, then governs all transitions in the transition energy bracket $[E^{B-}_j, E^{B+}_j]$. The corresponding polarization equation of motion for the bulk polarization density vector $\vec{P}^B_j(\mathbf{r},t)$ is presented in Eq.(2).

$$\frac{d^2 \vec{P}^B_j(\mathbf{r},t)}{dt^2} + \gamma^B_j \frac{d\vec{P}^B_j(\mathbf{r},t)}{dt} + \left[ \left(\omega^B_j\right)^2 + \frac{4\omega^2_{B_j} |\mu_{B_j}|^2}{\hbar^2} \vec{A}(\mathbf{r},t) \cdot \vec{A}(\mathbf{r},t) \right] \vec{P}_{B_j}(\mathbf{r},t)$$
$$= \frac{2\left(\omega^B_j\right)^2}{\hbar} |\mu^B_j|^2 \frac{1}{L_{ac}} \left[ N^B_{V(j)}(\mathbf{r},t) - N^B_{C(j)}(\mathbf{r},t) \right] \vec{E}(\mathbf{r},t) \qquad (2)$$

Equation (2) is similar to (1) but has different resonant frequencies $\omega^B_j$ which depend on the bulk band structure. Additionally, the dipole dephasing rates $\gamma^B_j$ are in general different from that of SQW. The light-matter interaction properties are isotropic as seen via the use of the same dipole matrix elements $|\mu^B_j|^2$ for all polarization density vector components. The RHS of Eq. (2) is dependent on the number of electrons per unit area $N^B_{V(j)}(\mathbf{r},t)$ in the $j^{th}$ valence band level and $N^B_{C(j)}(\mathbf{r},t)$ in the $j^{th}$ conduction band level.

In Eq. (2), only a single valence band is considered as the heavy-hole, light-hole degeneracy can be treated via a combined effective mass. The bulk heavy-hole and light-hole bands may also be treated separately, however the computational cost incurred in this case would be greater. While, it is necessary to consider the evolution of the carrier densities $N^B_{V(j)}(\mathbf{r},t)$ and $N^B_{C(j)}(\mathbf{r},t)$ for accurate calculation of the carrier dynamics, one may circumvent the calculation of Eq.(2) to improve computational efficiency. This is because, for most device applications, only light-matter interaction with the SQW's is important. However, in some cases, the bulk polarization density vector has been shown to be important for refractive index variations in SQW structures [30].

The various polarization density vectors in Eqs.(1)-(2) are coupled to the Maxwell's equations via the magnetic curl equation as shown in Section 8.

## 5. CALCULATION OF DIPOLE NUMBER DENSITIES

In this section, the calculation of dipole number density parameters $N_{dip(l,i)}^{qw}$ is presented. The specific details of the band structure are incorporated using this parameter.

### *5.1 SQW Dipole Number Density*

As mentioned in Section 4, the macroscopic polarization density vector $\vec{P}_{C-HH_{(l,i)}^{qw}}$ is the total number of microscopic dipole moments per unit volume in the energy bracket $[E_{C-HH(l,i)}^{qw-}, E_{C-HH(l,i)}^{qw+}]$ spanned by the dipole $\vec{\mu}_{C-HH(l,i)}^{qw}$. Since, dipoles are formed between electrons and holes, the dipole number density would depend on the number of available electronic states in the multi-electron broadened states $|l,i\rangle_C$ and $|l,i\rangle_{HH}$ (See Fig.3). The number of available electronic states can be calculated by using the density of states functions $g_C^{qw}$ in the conduction sub-band and $g_{HH}^{qw}$ in the heavy-hole sub-band. It is seen in Eq.(3) that as long as the $k_t$ conservation rule is satisfied (See Appendix A), the dipole number density $N_{dip(l,i)}^{qw}$ is equal to the total number of electron states per unit area in the conduction sub-band state $|l,i\rangle_C$ - $N_{C(l,i)}^{qw-0}$ (or total electron number density). This is in turn is equal to the total number of electronic states per unit area in the broadened heavy hole state $|l,i\rangle_{HH}$ - $N_{HH(l,i)}^{qw-0}$ (total heavy hole number density). In Eq.(3a), $E_{C(l,i)}^{qw-}$ and $E_{C(l,i)}^{qw+}$ represent the lower and upper bounds respectively of the energy bracket spanned by the broadened energy state $|l,i\rangle_C$ as shown in Fig.(3)( grey shaded regions, top figure). Similarly, in Eq.(3b), $E_{HH(l,i)}^{qw-}$ and $E_{HH(l,i)}^{qw+}$ represent the lower and upper bounds respectively of the energy bracket spanned by the broadened energy state $|l,i\rangle_{HH}$. When Eq.(3), is specifically evaluated for a SQW system with $n^{qw}$ wells and parabolic sub-bands, the dipole number density is given by Eq.(4).

$$N_{C(l,i)}^{qw-0}(\mathbf{r}) = \int_{E_{C(l,i)}^{qw-}}^{E_{C(l,i)}^{qw+}} g_C^{qw}(\mathbf{r},\varepsilon)d\varepsilon \tag{3a}$$

$$N_{HH(l,i)}^{qw-0}(r) = \int_{E_{HH(l,i)}^{qw-}}^{E_{HH(l,i)}^{qw+}} g_{HH}^{qw}(r,\varepsilon)d\varepsilon \tag{3b}$$

$$N_{dip(l,i)}^{qw}(\mathbf{r}) = N_{C(l,i)}^{qw-0}(\mathbf{r}) = N_{HH(l,i)}^{qw-0}(\mathbf{r}) \tag{3c}$$

$$N_{dip_{(l,i)}^{qw}}(\mathbf{r}) = n^{qw} \frac{m_{C(l)\|}^{qw} m_{HH(l)\|}^{qw}}{m_{C(l)\|}^{qw} + m_{HH(l)\|}^{qw}} \frac{1}{\pi\hbar^2} \Delta E_{C-HH(l,i)}^{qw} \tag{4}$$

Since an electron in a state $|l,i\rangle_C$ can make transitions to both the heavy-hole state $|l,i\rangle_{HH}$ as well as the light-hole state $|l,i\rangle_{LH}$, we also need to define a dipole number density parameter for conduction –light-hole transitions. It turns out that when the transverse momentum $k_t$ is conserved, the dipole number density for conduction to light-hole sub-band transitions is the same as that for conduction to heavy-hole sub-band transitions as shown in Eq.(5). Here, $N_{LH(l,i)}^{qw-0}(\mathbf{r})$ is called the total light-hole number density corresponding to state $|l,i\rangle_{LH}$.

$$N_{LH(l,i)}^{qw-0}(\mathbf{r}) = \int_{E_{HH(l,i)}^{qw-}}^{E_{HH(l,i)}^{qw+}} g_{LH}^{qw}(\mathbf{r},\varepsilon)d\varepsilon = N_{dip(l,i)}^{qw}(\mathbf{r}) \tag{5}$$

### 5.2. Bulk Dipole Number Density

The dipole number density $N_{dip(j)}^B(\mathbf{r})$ for dipoles $\vec{\mu}_j^B(\mathbf{r},t)$ governing transitions between bulk conduction and valence bands are also calculated similar to the SQW case. As in the SQW case, the dipole number density $N^B_{dip(j)}(\mathbf{r})$ is equal to the total number of electronic states per unit area in the conduction band state $|j\rangle_C$ - $N^{B-0}_{C(j)}(\mathbf{r})$. Furthermore, $N^{B-0}_{C(j)}(\mathbf{r})$ is in turn, equal to the total number of electronic states per unit area in the valence band state $|j\rangle_V$ - $N^{B-0}_{V(j)}(\mathbf{r})$. In [16]-[17], the dipole number density for bulk states was calculated in units of per unit volume. However, in the MB-MLME model, due to the effective medium treatment outlined in Section 3.3, we need to modify the calculation of the dipole number density into units of per unit area. This modification leads to the conservation of total sheet carrier density in both SQW and bulk states. Since, the barrier width is $L^B$, $N^B_{dip(j)}(\mathbf{r})$ is the total number of electronic states per unit volume in the energy bandwidth $[E^{B-}_{C(j)}, E^{B+}_{C(j)}]$ multiplied by the barrier width. Additionally, there are also bulk unconfined well states or SCH states which can be accounted for by multiplying the total number of states per unit volume by the by the corresponding widths of these regions. Then, in general, $N^B_{dip(j)}(\mathbf{r})$, $N^{B-0}_{C(j)}(\mathbf{r})$ and $N^{B-0}_{V(j)}(\mathbf{r})$ may be calculated as in Eq.(6).

$$N^{B-0}_{V(j)}(\mathbf{r}) = \int_{E^{B-}_{C(j)}}^{E^{B+}_{C(j)}} \left(n^{qw}+1\right) L^B g^{barr}_C(\mathbf{r},\varepsilon) d\varepsilon + \int_{E^{B-}_{C(j)}}^{E^{B+}_{C(j)}} n^{qw} L^{qw} g^{un}_C(\mathbf{r},\varepsilon) d\varepsilon$$
$$+ \int_{E^{B-}_{C(j)}}^{E^{B+}_{C(j)}} L^{SCH} g^{SCH}_C(\mathbf{r},\varepsilon) d\varepsilon \tag{6a}$$

In Eq.(6a), $g^{barr}_C$, $g^{un}_C$ and $g^{SCH}_C$ correspond to the bulk density of states functions in the barrier, unconfined well and SCH regions respectively. The first term in Eq.(6a) contains a factor of $n^{qw}+1$ to consider the total number of states in all the barriers.

$$N^B_{dip(j)}(r) = N^{B-0}_{V(j)}(\mathbf{r}) = N^{B-0}_{C(j)}(\mathbf{r}) \tag{6b}$$

If one uses an effective mass approximation for the bulk bandstructure, all the bulk regions(barrier+unconfined well+SCH) may be represented by a single effective mass

$m_C^B$ corresponding to the sum of the bulk density of states of all regions as shown in Eq.(7). A similar expression can be used for a combined valence band effective mass $m_V^B$.

$$m_C^B = \left( \frac{(n^{qw}+1)L^b \left(m_C^{barr}\right)^{3/2} + n^{qw}L^{qw}\left(m_C^{un}\right)^{3/2} + L^{SCH}\left(m_C^{SCH}\right)^{3/2}}{(n^{qw}+1)L^b + n^{qw}L^{qw} + L^{SCH}} \right)^{2/3} \quad (7)$$

Since our approach to thermalizing carriers based on EUT and EDT outlined in Section 3 to the correct Fermi-Dirac distributions only depends on the parameters $N_{C(j)}^{B-0}(\mathbf{r})$ and $N_{V(j)}^{B-0}(\mathbf{r})$ (See Section 7), this simple modification of using a combined effective mass in Eq.(7), will ensure that the carriers are thermalized to the correct Fermi-Dirac distributions.

## 6. Carrier rate equations

In, Eqs.(1)-(2),the polarization density vectors were dependent on the carrier number densities of each broadened state. In this section, we present carrier rate equations for carrier populations in the SQW states as well as bulk states.

### *6.1. SQW carrier rate equations*

In general, carriers in the SQW sub-bands can make three types of transitions (1) Interband transitions between conduction and heavy-hole or conduction and light-hole sub-bands. (2) Intra sub-band transitions which occur within each conduction, heavy-hole or light-hole sub-band and (3) Inter sub-band transitions between any two conduction or valence (heavy and light-hole) sub-bands. Then, the overall carrier rate equation for the electron density $N_{C(l,i)}^{qw}(\mathbf{r},t)$ in the state $|l,i\rangle_C$ is presented in Eq.(8).

$$\frac{dN_{C(l,i)}^{qw}(\mathbf{r},t)}{dt} = \Delta N_{C-HH(l,i)}^{qw} + \Delta N_{C-LH(l,i)}^{qw} + \Delta N_{C(l,i)}^{qw} + \Delta N_{C(l)-C}^{qw} + \Delta N_{C(l)-B} \quad (8)$$

In Eq.(8), the first two terms represent the rate of interband transitions. The first term corresponds to transitions between conduction and heavy-hole sub-bands. While, the second term corresponds to conduction-light hole transitions. The rate of conduction to heavy-hole transitions are denoted by $\Delta N^{qw}_{C-HH(l,i)}$ to indicate the change in carrier population that occurs in a small time interval $\delta t$. The subscript here indicates that transitions occur between the $i^{th}$ level of the $l^{th}$ conduction sub-band and the $i^{th}$ level of the $l^{th}$ heavy-hole sub-band. Thus these transitions satisfy the $\Delta l=0$ and $\Delta k_t=0$ rules. Similarly, the rate of conduction to light-hole transitions are denoted by $\Delta N^{qw}_{C-LH(l,i)}$. The third term represents the rate of intra sub-band transitions for the conduction sub-band. In our convention, intra sub-band transitions are denoted by $\Delta N^{qw}_{\alpha(l,i)}$, where $\alpha$ is a generic symbol for conduction (C), heavy-hole (HH) or light-hole (LH) sub-bands. The fourth term $\Delta N^{qw}_{C(l)-C}$ in Eq.(8) represents the rate of inter sub-band transitions. The subscript '$C(l)-C$' indicates that transitions between the $l^{th}$ conduction sub-band and all other conduction sub-bands are considered. Since the inter sub-band terms and intra sub-band terms occur only within the SQW states and cause no overall change to the SQW carrier density. Their aggregate over all states would be zero. The final term $\Delta N_{C(l)-B}$ represents the rate of transitions between conduction band states in the SQW and barrier. The subscript '$C(l)-B$' indicates that transitions occur between the $l^{th}$ SQW conduction sub-band and the bulk conduction band. Similar to Eq.(8), carrier rate equations for the electron densities $N^{qw}_{HH(l,i)}$ in the heavy hole states $|l,i\rangle_{HH}$ and electron densities $N^{qw}_{LH(l,i)}$ in the light hole states $|l,i\rangle_{LH}$ are presented in Eqs.(9) and (10) respectively.

$$\frac{dN^{qw}_{HH(l,i)}(\mathbf{r},t)}{dt} = -\Delta N^{qw}_{C-HH(l,i)} + \Delta N^{qw}_{HH(l,i)} + \Delta N^{qw}_{HH(l)-HH} + \Delta N^{qw}_{HH(l)-LH} + \Delta N_{HH(l)-B} \qquad (9)$$

$$\frac{dN_{LH(l,i)}^{qw}(\mathbf{r},t)}{dt} = -\Delta N_{C-LH(l,i)}^{qw} + \Delta N_{LH(l,i)}^{qw} + \Delta N_{LH(l)-LH}^{qw} + \Delta N_{LH(l)-HH}^{qw} + \Delta N_{LH(l)-B} \qquad (10)$$

The various terms in Eqs.(9) and (10) have similar meanings as that in Eq.(8). The first term in Eqs.(9) and (10) represent interband transitions. Note that the signs of the first terms in Eqs.(9) and (10) are opposite of those in Eq.(8). This is because, electrons lost in these valence band states are gained in the corresponding conduction band states and vice versa. The second terms in Eqs.(9) and (10) correspond to the intra sub-band transition terms and will be discussed in Section 7. The third terms $\Delta N_{HH(l)-HH}^{qw}$ and $\Delta N_{LH(l)-LH}^{qw}$ in Eqs.(9) and (10) are similar to the $\Delta N_{C(l)-C}^{qw}$ terms described previously. For the case of the heavy-hole or light-hole bands, inter sub-band transitions also occur between heavy-holes and light-holes. Thus, $\Delta N_{LH(l)-HH}^{qw}$ represents the rate of transitions between the $l^{th}$ light-hole sub-band and all heavy-hole sub-bands. Similarly, $\Delta N_{HH(l)-LH}^{qw}$ represents the rate of transitions between the $l^{th}$ heavy-hole and all other light-hole sub-bands. The final terms in Eqs.(9) and (10) represent the rate of transitions to the barrier states, similar to the terms $\Delta N_{C(l)-B}$ in Eq.(8).

In Eq.(11), we present the interband transition rates $\Delta N_{C-HH(l,i)}^{qw}$ for conduction to heavy-hole transitions.

$$\Delta N_{C-HH(l,i)}^{qw} = -\frac{\omega_{C-HH(l,i)}^{qw}}{\hbar}\vec{A}(\mathbf{r},t).\vec{P}_{C-HH(l,i)}^{qw}(\mathbf{r},t) - \frac{N_{C(l,i)}^{qw}(\mathbf{r},t)}{\tau_{C-HH(l,i)}^{qw}}\left(1 - \frac{N_{HH(l,i)}^{qw}(\mathbf{r},t)}{N_{HH(l,i)}^{qw-0}(\mathbf{r})}\right) \qquad (11)$$

An electron in the heavy-hole band can make a transition to the conduction band by stimulated absorption or an electron in the conduction band can make a transition to the heavy-hole band by stimulated emission. This is captured by the first term in Eq.(11). Here $\vec{A}(\mathbf{r},t)$ is the electric vector potential and $\vec{P}_{C-HH(l,i)}^{qw}(\mathbf{r},t)$ is the polarization density vector whose dynamics are described by Eq.(1). The second term in Eq.(11) represents the spontaneous decay of an electron in the conduction band to the heavy hole band with a transition time $\tau_{C-HH(l,i)}^{qw}$. The factor $\left(1 - N_{HH(l,i)}^{qw}(\mathbf{r},t) / N_{HH(l,i)}^{qw-0}(\mathbf{r})\right)$ in the second term of Eq.(11) represents the Pauli's exclusion principle and represents the fact that an electron in the conduction band state can decay to the heavy-hole band only if the corresponding heavy-hole state is vacant. Similar to Eq.(11), we can write the conduction to light- hole transitions by replacing the subscripts/superscripts in Eq.(11) 'HH' with 'LH'.

### *6.2. Bulk carrier rate equations*

In this section, we track the carrier dynamics of the bulk states which include the bulk, barrier, unconfined well and even SCH states. In Eq.(12), we present the overall rate equation for the conduction band electrons in the bulk states.

$$\frac{dN_{C(j)}^{B}(\mathbf{r},t)}{dt} = \Delta N_{j}^{B} + \Delta N_{C(j)}^{B} - \sum_{l} \Delta N_{C(l)-B} \qquad (12)$$

The first term of Eq.(12) represents interband transitions. The second term represents intra-band transitions which thermalizes carriers to the correct Fermi-Dirac distributions. The final term represents the sum of all electronic transitions between various SQW conduction sub-bands and the bulk states presented in the final term of Eq.(8). Note that the sign of the third term is opposite to that of Eq.(8). In fact upon adding the rate of carrier exchange between all SQW sub-bands and the bulk states, it is seen that the net exchange of carriers would be zero. This is because any carrier escaping from the SQW states is captured by the bulk states and vice versa.

Similar to Eq.(12), we can write the rate equation for valence band electrons in the bulk states according to Eq.(13).

$$\frac{dN_{V(j)}^{B}(\mathbf{r},t)}{dt} = -\Delta N_{j}^{B} + \Delta N_{V(j)}^{B} - \sum_{l} \Delta N_{HH(l)-B} - \sum_{m} \Delta N_{LH(m)-B} \qquad (13)$$

Note that, upon adding Eqs.(8)-(10) and (12-(13), the total rate of change is always zero. Thus, the total sheet carrier density is always conserved. The explicit form of the interband transition terms $\Delta N_j^B$ in Eq.(12) and (13) is presented in Eq.(14).

$$\Delta N_j^B = -\frac{\omega_j^B}{\hbar}\vec{A}(\mathbf{r},t).\vec{P}_j^B(\mathbf{r},t) - \frac{N_{C(j)}^B(\mathbf{r},t)}{\tau_{B(j)}}\left(1 - \frac{N_{V(j)}^B(\mathbf{r},t)}{N_{V(j)}^{B-0}(\mathbf{r})}\right) + R_{pump}\left(\frac{N_{V(j)}^B(\mathbf{r},t)}{N_{V(j)}^{B-0}(\mathbf{r})}\right)\left(1 - \frac{N_{C(j)}^B(\mathbf{r},t)}{N_{C(j)}^{B-0}(\mathbf{r})}\right)\delta_{j,j_{max}}$$

(14)

While the first two terms represent stimulated emission/absorption and spontaneous decay, the third term represents electrical pumping. The electrical pumping terms are present only for the highest bulk state level as indicated by the Dirac-Delta factor $\delta_{j,j_{max}}$. The electrons make their way to the SQW through the carrier leakage terms in Eq.(12) after being injected into the bulk states. Furthermore, electrical pumping can occur only if there are electrons in the valence band state and vacancies in the conduction band state as delineated by the pump blocking factor.

### *6.3. Including Auger Recombination*

In SQW's, Auger recombination losses can be significant [31]. Previously, in our work on the MLME model [16]-[17], Auger recombination was not included. While a microscopic approach to modelling Auger recombination can be quite cumbersome, we include the effect phenomenologically via the typical Auger recombination coefficient $C^{qw}$. In our model, we modify the electrical $R_{pump}$ term in Eq. (14) according to Eq.(15) by subtracting the total Auger recombination rate (per unit area) from the injection current density. The Auger recombination rate (per unit volume) per well is proportional to the cube of the total carrier density per well. Therefore, we calculate the total carrier density by summing the sheet carrier densities $N_{C(l,i)}^{qw}(\mathbf{r},t)$ over all the energy levels $i$ and SQW sub-bands $l$ and dividing it by the total SQW width as seen in the denominator of Eq.(15).

$$R_{pump} = \frac{J}{q} - n^{qw}L^{qw}C^{qw}\left(\frac{\sum_l \sum_i N^{qw}_{C(l,i)}(\mathbf{r},t)}{n^{qw}L^{qw}}\right)^3 \quad (15)$$

The total Auger recombination rate (per unit area) for all wells will then be multiplied by a pre-factor $n^{qw}L^{qw}C^{qw}$. Here, the pumping rate $R_{pump}$ is written in units of per unit area per second and hence the current injection is written as $q^{-1}J$ rather than the usual $(qd)^{-1}J$ [16], where $d$ is the active region thickness. In our current approach the effect of the active region thickness is indirectly included via the incorporation of barrier/well widths in the calculation of the electron number density parameters in Eq.(3).

## 7. Thermalizing carrier distributions in the MB-MLME model

In this section, we present the key formulations for the various intra sub-band, inter sub-band and carrier leakage terms. These terms are critical to the computational efficiency of the MB-MLME model as they circumvent numerically cumbersome iterative procedures. As outlined in Section 3.2, these terms thermalize the carrier distributions in the SQW and bulk states to a common Fermi-Dirac function at quasi-equilibrium. In Fig.4 and Fig.5, we presented the scheme of these transitions using the concept of EDT and EUT. In this section, we present the corresponding mathematical equations. Similar to [16]-[17], EDT occurs with transition time $\tau_d$ and EUT with transition time $\tau_u$ between two states. The carrier distribution would be thermalized to the correct Fermi-Dirac distribution if $\tau_u$ is related to $\tau_d$ according to Eq.(16) and if Pauli's exclusion is included (See Appendix B for derivation).

$$\tau_u = \tau_d \frac{N_d^0}{N_u^0} \exp\left\{\frac{E_u - E_d}{k_B T_L}\right\} \qquad (16)$$

In Eq.(16), $T_L$ is the lattice temperature; $E_u$ is the energy of the upper state and $E_d$ is the energy of the lower state. $N_d^0$ and $N_u^0$ are the total number of electronic states per unit area in the broadened lower state and upper state respectively. Therefore, the approach of using EUT and EDT is valid in general for many band structures

Previously, in [16]-[17] it was shown that if the intraband transition terms are written so as to conserve the total number of carriers per unit volume, the correct Fermi-Dirac distributions are obtained for bulk semiconductors. In this paper, since the carrier rate equations are formulated so as to conserve the total sheet carrier density (number of carriers per unit area), carriers from various SQW and bulk states are thermalized to a common Fermi-Dirac distribution.

### 7.1. Intra sub-band transitions

Let us consider the thermalization of carriers within the $l^{th}$ conduction sub-band. The scheme of transitions is depicted in Fig.4a. An electron in the $i^{th}$ level of the $l^{th}$ conduction sub-band can make EDT with a transition time $\tau_d = \tau_{[l,(i,i-1)]C}$ to the $(i-1)^{th}$ level in the same sub-band. Of course, this is provided there is no electron in this level. Simultaneously, an electron in the $(i-1)^{th}$ level can make an EUT with a transition time $\tau_u = \tau_{[l,(i-1,i)]C}$ to the $i^{th}$ level as dictated by Pauli's exclusion principle. Similarly, EUT and EDT can also occur between the $i^{th}$ and $(i+1)^{th}$ levels. Thus, we write the total intra sub-band transition rate $\Delta N_{\alpha(l,i)}^{qw}$ for the $i^{th}$ level, in the $l^{th}$ conduction sub-band according to Eq.(17).

$$\Delta N_{C(l,i)}^{qw} = \Delta N_{[l,(i+1,i)]C} - \Delta N_{[l,(i,i-1)]C} \qquad (17a)$$

$$\Delta N_{[l,(i,i-1)]C} = \frac{N_{C(l,i)}^{qw}(\mathbf{r},t)}{\tau_{[l,(i,i-1)]C}}\left(1-\frac{N_{C(l,i-1)}^{qw}(\mathbf{r},t)}{N_{C(l,i-1)}^{qw-0}(\mathbf{r})}\right) - \frac{N_{C(l,i-1)}^{qw}(\mathbf{r},t)}{\tau_{[l,(i-1,i)]C}}\left(1-\frac{N_{C(l,i)}^{qw}(\mathbf{r},t)}{N_{C(l,i)}^{qw-0}(\mathbf{r})}\right) \quad (17b)$$

In reality, one has to consider transitions between any two levels *(i,j)* in the $l^{th}$ conduction sub-band. However, EUT times are exponentially dependent on the energy separation between the two states involved. Since, the energy separation between two levels in the $l^{th}$ conduction sub-band $\Delta E_{C(l,i)}^{qw} \approx \hbar \gamma_{C-HH(l,i)}^{qw} \approx k_B T_L$, only transitions between adjacent levels are most significant. In Eq.(17), the subscripts '[*l*,(i-1,i)]C' are used to denote that transitions occur from the *(i-1)*$^{th}$ level to the $i^{th}$ level in the $l^{th}$ conduction sub-band. In Eq.17a, the first term denotes the increase in the number of electrons per unit area due to net downward transition of electrons to the $i^{th}$ level from the *(i+1)*$^{th}$ level in the $l^{th}$ sub-band. Similarly, the second term represents the loss of electrons due to the net downward transition from the $i^{th}$ to the *(i-1)*$^{th}$ levels within the same sub-band. The general definition of the $\Delta N_{[l,(i,i-1)]C}$ terms is given in Eq.17b. Here the carrier transitions are constrained by the Pauli's exclusion principle as described above. Using the general relation in Eq. (16), we have the relation between $\tau_{[l,(i,i-1)]C}$ and $\tau_{[l,(i-1,i)]C}$ in Eq.(18)

$$\tau_{[l,(i-1,i)]C} = \tau_{[l,(i,i-1)]C}\frac{N_{C(l,i-1)}^{qw-0}}{N_{C(l,i)}^{qw-0}}\exp\left\{\frac{E_{C(l,i)}^{qw} - E_{C(l,i-1)}^{qw}}{k_B T_L}\right\} \quad (18)$$

In Eq.(18), $E_{C(l,i)}^{qw}$ is the energy of the $i^{th}$ level of the $l^{th}$ conduction sub-band and $N_{C(l,i)}^{qw-0}$ is the corresponding total density of electronic states per unit area as defined in Eq.(3). The intra sub-band transition rates for heavy-hole sub-bands $\Delta N_{HH(l,i)}^{qw}$ and light-hole sub-bands $\Delta N_{LH(l,i)}^{qw}$ can be written similarly. Thus Eqs.(17) and (18) will drive electrons in the $l^{th}$ conduction sub-band to a Fermi-Dirac distribution.

## 7.2. Inter sub-band transitions

Intra sub-band transitions only thermalize carriers within each sub-band. In order to thermalize carriers in the various sub-bands to a common Fermi-Dirac distribution, a scheme of inter sub-band transitions was outlined in Fig.4b. In Fig.4b, it can be seen that EDT and EUT occur between the bottom of any two conduction sub-bands $l$ and $m$. In general, transitions can occur between any two conduction sub-band states $|l,i\rangle_C, |m,j\rangle_C$. However, we restrict ourselves to a simplified scheme of transitions between the bottom most states $|l,1\rangle_C, |m,1\rangle_C$ of sub-bands $l,m$. This is because the zone centers of the sub-bands are strongly coupled by electron-phonon scattering processes [23]-[24].

An electron in the bottom most energy level of the $l^{th}$ conduction sub-band can make downward transitions with transition time $\tau_d = \tau_{[(l,m),1]C}$ to the bottom most energy level of every $m^{th}$ conduction sub-band (for $m<l$). Here the subscript '[(l,m),1]C' denotes a transition between the first states in $l^{th}$ and $m^{th}$ sub-bands. Similarly an electron from every $m^{th}$ sub-band can make EUT with a transition time $\tau_u = \tau_{[(m,l),1]C}$ to the $l^{th}$ sub-band. Similarly, the electron in the $l^{th}$ conduction sub-band can make EUT with transition time $\tau_u = \tau_{[(l,m),1]C}$ to every $m^{th}$ sub-band ($m>l$) and electrons in these sub-bands can make EDT with transition time $\tau_d = \tau_{[(m,l),1]C}$ to the $l^{th}$ sub-band. Thus, the total rate of inter sub-band transitions in the conduction sub-band $l$ $\Delta N_{C(l)-C}$ is given by Eq.(19).

$$\Delta N_{C(l)-C} = \sum_{m>l} \Delta N_{[(m,l),1]C} - \sum_{m<l} \Delta N_{[(l,m),1]C} \qquad (19a)$$

$$\Delta N_{[(l,m),1]C} = \frac{N^{qw}_{C(l,1)}(\mathbf{r},t)}{\tau_{[(l,m),1]C}} \left(1 - \frac{N^{qw}_{C(m,1)}(\mathbf{r},t)}{N^{qw-0}_{C(m,1)}(\mathbf{r})}\right) \\ - \frac{N^{qw}_{C(m,1)}(\mathbf{r},t)}{\tau_{[(m,l),1]C}} \left(1 - \frac{N^{qw}_{C(l,1)}(\mathbf{r},t)}{N^{qw-0}_{C(l,1)}(\mathbf{r})}\right) \qquad (19b)$$

In Eq.(19a), the first term indicates the increase in the number of electrons per unit area due to the net downward transition of electrons to the $l^{th}$ sub-band due to all other sub-bands $m>l$. Similarly, the second term in Eq.(19a) represents the net loss of electrons per unit area due to downward transition of electrons from the sub-band $l$ to every other sub-band $m<l$. The definition of the $\Delta N_{[(l,m),1]C}$ terms in Eq. (19a) are provided in Eq.(19b). Equation 19(b) contains the exchange of electrons between sub-bands $l$ and $m$ and obeys Pauli's exclusion principle as previously discussed. Using the relation between EUT and EDT times given in Eq.(16), the ratio of upward and downward inter sub-band transition times are given by Eq.(20).

$$\tau_{[(m,l),1]C} = \tau_{[(l,m),1]C} \frac{N_{C(m,1)}^{qw-0}}{N_{C(l,1)}^{qw-0}} \exp\left\{\frac{E_{C(l,1)}^{qw} - E_{C(m,1)}^{qw}}{k_B T_L}\right\} \forall (l>m) \tag{20}$$

Equations (19) and (20) will tend to pin the bottom most energy levels of sub-bands $m$ and $l$ to the same Fermi-Dirac distribution. Simultaneously, the intra sub-band terms $\Delta N_{C(l,i)}^{qw}$ (Eq.17) for the $l^{th}$ sub-band will tend to drive all the carriers within that sub-band to the same Fermi-Dirac distribution. Similarly, the intra sub-band terms $\Delta N_{C(m,i)}^{qw}$ will tend to drive the carriers within the $m^{th}$ sub-band to a common Fermi-Dirac distribution. Thus all the sub-bands will converge towards a common Fermi-Dirac distribution over a time scale given proportional to $\tau_{[(m,l),1]C}$. All other inter sub-band transition rates- $\Delta N_{HH(l)-HH}^{qw}$, $\Delta N_{HH(l)-LH}^{qw}$ and $\Delta N_{LH(l)-LH}^{qw}$ can be written similarly.

### *7.3. Modeling leakage to the bulk states*

In this section, we further include carrier leakage terms $\Delta N_{C(l)-B}$ to account for carrier leakage from SQW's to the bulk states. Since, we can include all barrier, unconfined as well as SCH states via the density of state parameters in Eq.(7), these $\Delta N_{C(l)-B}$ terms effectively thermalize carriers between the SQW and all these states. The scheme for these leakage terms is presented in Fig.5. In Fig.5, we see that these EUT and EDT terms are applied between the bulk and SQW states in our effective medium approach. It is tantamount to repeated carrier capture and escape processes. A drawback with this effective medium approach however is that it does not account for spatial inhomogeneity of carrier distribution within the multiple well-barrier system due to tunneling processes. However, these tunneling terms may be included by a separate rate equation as in [14],[32]. In Fig.5, we see that transitions occur between the bottom of the bulk states and the bottom of the various SQW sub-bands. As in Section 7.2, only transitions between the bottom of the bulk band and SQW sub-bands are considered. In reality, the capture and escape process is more complicated. A detailed treatment would require accounting for all possible scattering processes which is not computationally efficient for FDTD simulations. The formulation for $\Delta N_{C(l)-B}$ follows on the same lines as the intra sub-band and inter sub-band transitions in Sections 7.1 and 7.2. Thus, we directly present the results in Eqs.(21) and (22). The corresponding leakage terms for the valence band states $\Delta N_{HH(l)-B}$ and $\Delta N_{LH(l)-B}$ are formulated similarly.

$$\Delta N_{C(l)-B} = \frac{N_{C(1)}^{B}}{\tau_{B \to C(l)}}\left(1 - \frac{N_{C(l,1)}^{qw}}{N_{C(l,1)}^{qw-0}}\right) - \frac{N_{C(l,1)}^{qw}}{\tau_{C(l) \to B}}\left(1 - \frac{N_{C(1)}^{B}}{N_{C(1)}^{B-0}}\right) \tag{21}$$

$$\tau_{C(l) \to B} = \tau_{B \to C(l)} \frac{N_{C(l,1)}^{qw-0}}{N_{C(1)}^{B-0}} \exp\left\{\frac{E_{C(1)}^{B} - E_{C(l,1)}^{qw}}{k_B T_L}\right\} \tag{22}$$

Equations (21) and (22) will drive the bulk states and the $l^{th}$ conduction sub-band towards the same Fermi-Dirac distribution. However, all the sub-bands are coupled by the inter sub-band transitions in Section 7.2. Therefore, the bulk and SQW states converge to a common Fermi-Dirac distribution.

## 8. Coupling of medium equations to Maxwell's Equations

We couple the polarization density vectors in Section 4 to the magnetic curl equation of the Maxwell's equations as shown in Eq. (23). We include the confinement factor which is required for 1-D or 2-D simulations. For the general 3-D case, this is not required. Note that $\Gamma$ here is the confinement factor for the complete active region. The individual overlap with the barriers and SQW automatically manifests itself due to the formulation in terms of sheet carrier density. The final term in Eq.(23) incorporates free carrier absorption and intervalence band absorption via the conductivity terms $\beta_{FCA}, \beta_{IV}$ respectively. Since the polarization terms are dependent on the carrier dynamics. Eqs.(1)-(23) are all coupled to each other in space and time.

$$\nabla \times \vec{H}(\mathbf{r},t) = \varepsilon_b(\mathbf{r}) \frac{d\vec{E}(\mathbf{r},t)}{dt} + \Gamma \left( \begin{array}{l} \sum_l \sum_i \frac{\partial \vec{P}^{qw}_{C-HH(l,i)}(\mathbf{r},t)}{\partial t} \\ + \sum_m \sum_i \frac{\partial \vec{P}^{qw}_{C-LH(m,i)}(\mathbf{r},t)}{\partial t} + \sum_j \frac{\partial \vec{P}^{B}_{j}(\mathbf{r},t)}{\partial t} \end{array} \right) \quad (23)$$
$$+ (\beta_{IV} + \beta_{FCA}) \vec{E}(\mathbf{r},t)$$

## 9. FDTD update equations

In this section, we present the update equations for the implementation of the MB-MLME FDTD model. For quick reference of the various MB-MLME parameters, we once again point the reader to Appendix C which lists all parameters along with their definitions. The FDTD update equations are written according to the Yee Algorithm [1] for a 3-D Cartesian space. The discrete variables $(u,v,w)$ are used to denote the spatial co-ordinates $(x,y,z) = (u\Delta x, v\Delta y, w\Delta z)$, where $\Delta x, \Delta y, \Delta z$ are the spatial resolutions. The discrete variable $n$ is used to denote the time instant $t = n\Delta t$, where $\Delta t$ is the temporal resolution. The perfected matched layer boundary conditions are implemented using the UPML algorithm [33] and field sources are implemented using the total field/scattered field (TF/SF) method [34].

In Eq.(24), we present the update equation corresponding to the polarization equation of motion governing transitions between the conduction and heavy hole sub-bands from Eq.(1). The dipoles, carrier populations are centered at the same grid location as the electric fields but the magnetic fields are staggered by half a grid [1]. Equation (24) presents the update equation for the $z$ component, with other components updated similarly. Furthermore, similar update equations are written for other polarization density vectors.

$$P_{z[C-HH(l,i)]}^{qw}\Big|_{u-1/2,v+1/2,w}^{n+1} = \frac{4 - 2\Delta t^2 \left( \left(\omega_{C-HH(l,i)}^{qw}\right)^2 + 4\frac{\left(\omega_{C-HH(l,i)}^{qw}\right)^2}{\hbar^2} \left|\mu_{C-HH(l,i)}^{qw}\right|^2 A_z^2\Big|_{u-1/2,v+1/2,w}^{n} \right)}{2 + \Delta t \gamma_{C-HH(l,i)}^{qw}} P_{z[C-HH(l,i)]}^{qw}\Big|_{u-1/2,v+1/2,w}^{n}$$
$$+ \frac{\Delta t \gamma_{C-HH(l,i)}^{qw} - 2}{\Delta t \gamma_{C-HH(l,i)}^{qw} + 2} P_{z[C-HH(l,i)]}^{qw}\Big|_{u-1/2,v+1/2,w}^{n-1} - \frac{4\Delta t^2}{\hbar \left(\Delta t \gamma_{C-HH(l,i)}^{qw} + 2\right)} \left|\mu_{C-HH(l,i)}^{qw}\right|^2 \left( N_{C(l,i)}^{qw}\Big|_{u-1/2,v+1/2,w}^{n} - N_{HH(l,i)}^{qw}\Big|_{u-1/2,v+1/2,w}^{n} \right) E_z\Big|_{u-1/2,v+1/2,w}^{n}$$
(24)

Next, the electric and magnetic fields are updated according to (25) corresponding to the curl equation presented in (23).

$$E_z\Big|_{u-1/2,v+1/2,w}^{n+1} = E_z\Big|_{u-1/2,v+1/2,w}^{n}$$
$$- \frac{\Gamma}{\varepsilon_b\Big|_{u-1/2,v+1/2,w}} \sum_l \sum_i \left( P_{z[C-HH(l,i)]}^{qw}\Big|_{u-1/2,v+1/2,w}^{n+1} - P_{z[C-HH(l,i)]}^{qw}\Big|_{u-1/2,v+1/2,w}^{n} \right)$$
$$- \frac{\Gamma}{\varepsilon_b\Big|_{u-1/2,v+1/2,w}} \sum_l \sum_i \left( P_{z[C-HH(l,i)]}^{qw}\Big|_{u-1/2,v+1/2,w}^{n+1} - P_{z[C-HH(l,i)]}^{qw}\Big|_{u-1/2,v+1/2,w}^{n} \right)$$
$$- \frac{\Gamma}{\varepsilon_b\Big|_{u-1/2,v+1/2,w}} \sum_j \left( P_{z(j)}^{B}\Big|_{u-1/2,v+1/2,w}^{n+1} - P_{z(j)}^{B}\Big|_{u-1/2,v+1/2,w}^{n} \right)$$
$$+ \frac{\Delta t}{\varepsilon_b\Big|_{u-1/2,v+1/2,w}} \left( \frac{H_y\Big|_{u,v+1/2,w}^{n+1/2} - H_y\Big|_{u-1,v+1/2,w}^{n+1/2}}{\Delta x} - \frac{H_x\Big|_{u-1/2,v+1,w}^{n+1/2} - H_x\Big|_{u-1/2,v,w}^{n+1/2}}{\Delta y} \right)$$
$$- \Delta t (\beta_{IV} + \beta_{FCA}) E_z\Big|_{u-1/2,v+1/2,w}^{n}$$
(25)

The updated electric field is then used to update the electric vector potential according to Eq.(26).

$$A_z\Big|_{u-1/2,v+1/2,w}^{n+1} = A_z\Big|_{u-1/2,v+1/2,w}^{n} - \frac{\Delta t}{2}\left( E_z\Big|_{u-1/2,v+1/2,w}^{n+1} + E_z\Big|_{u-1/2,v+1/2,w}^{n} \right)$$
(26)

The temporal update of the carrier population variables can be performed with a larger time step $\kappa\Delta t$ $(\kappa \gg 1)$ in comparison to the electromagnetic variables as they evolve over time scales two orders of magnitude larger than the fields [35]. This approach has been shown to reduce computational time by > 3 times when $\kappa \sim 100$[35]. Thus, the discrete variable $\kappa n$ is used to denote that carrier populations are updated every $\kappa$ time steps.

In Eq.(27), we present the update equation for the interband transitions between conduction and heavy-hole sub-bands based on Eq.(11).

$$\Delta N^{qw}_{C-HH(l,i)}\Big|^{\kappa n}_{u-\frac{1}{2},v+\frac{1}{2},w} = -\frac{\omega^{qw}_{C-HH(l,i)}}{\hbar}\vec{A}\Big|^{\kappa n}_{u-\frac{1}{2},v+\frac{1}{2},w}\cdot\vec{P}^{qw}_{C-HH(l,i)}\Big|^{\kappa n}_{u-\frac{1}{2},v+\frac{1}{2},w}$$
$$-\frac{N^{qw}_{C(l,i)}\Big|^{\kappa n}_{u-\frac{1}{2},v+\frac{1}{2},w}}{\tau^{qw}_{C-HH(l,i)}}\left(1-\frac{N^{qw}_{HH(l,i)}\Big|^{\kappa n}_{u-\frac{1}{2},v+\frac{1}{2},w}}{N^{qw-0}_{HH(l,i)}\Big|_{u-\frac{1}{2},v+\frac{1}{2},w}}\right) \quad (27)$$

Interband transition rates $\Delta N^{qw}_{C-LH(l,i)}$ are updated similarly. In addition to the interband transitions, one has to account for intra sub-band transitions within each conduction sub-band. The update equations for these processes are presented in Eq.(28) based on Eq.(17).

$$\Delta N^{qw}_{C(l,i)}\Big|^{\kappa n}_{u-\frac{1}{2},v+\frac{1}{2},w} = \Delta N_{[l,(i+1,i)]C}\Big|^{\kappa n}_{u-\frac{1}{2},v+\frac{1}{2},w} - \Delta N_{[l,(i,i-1)]C}\Big|^{\kappa n}_{u-\frac{1}{2},v+\frac{1}{2},w} \quad (28a)$$

$$\Delta N_{[l,(i,i-1)]C}\Big|^{\kappa n}_{u-\frac{1}{2},v+\frac{1}{2},w} = \frac{N^{qw}_{C(l,i)}\Big|^{\kappa n}_{u-\frac{1}{2},v+\frac{1}{2},w}}{\tau_{[l,(i,i-1)]C}}\left(1-\frac{N^{qw}_{C(l,i-1)}\Big|^{\kappa n}_{u-\frac{1}{2},v+\frac{1}{2},w}}{N^{qw-0}_{C(l,i-1)}\Big|_{u-\frac{1}{2},v+\frac{1}{2},w}}\right)$$
$$-\frac{N^{qw}_{C(l,i-1)}\Big|^{\kappa n}_{u-\frac{1}{2},v+\frac{1}{2},w}}{\tau_{[l,(i-1,i)]C}}\left(1-\frac{N^{qw}_{C(l,i)}\Big|^{\kappa n}_{u-\frac{1}{2},v+\frac{1}{2},w}}{N^{qw-0}_{C(l,i)}\Big|_{u-\frac{1}{2},v+\frac{1}{2},w}}\right) \quad (28b)$$

Similarly, inter sub-band transitions from Eq.(19) and carrier leakage terms from Eq.(21) are updated. All transitions are then combined to provide the overall update of the electron density of the $i^{th}$ level in the $l^{th}$ conduction sub-band as shown in Eq.(29) which corresponds to Eq.(8).

$$N_{C(l,i)}^{qw}\Big|_{u-\frac{1}{2},v+\frac{1}{2},w}^{(\kappa+1)n} = \kappa\Delta t \begin{pmatrix} \Delta N_{C-HH(l,i)}^{qw}\Big|_{u-\frac{1}{2},v+\frac{1}{2},w}^{\kappa n} + \Delta N_{C-LH(l,i)}^{qw}\Big|_{u-\frac{1}{2},v+\frac{1}{2},w}^{\kappa n} \\ + \Delta N_{C(l,i)}^{qw}\Big|_{u-\frac{1}{2},v+\frac{1}{2},w}^{\kappa n} \\ + \Delta N_{C(l)-C}^{qw}\Big|_{u-\frac{1}{2},v+\frac{1}{2},w}^{\kappa n} + \Delta N_{C(l)-B}\Big|_{u-\frac{1}{2},v+\frac{1}{2},w}^{\kappa n} \end{pmatrix} \quad (29)$$

Similar update equations are written for the heavy-hole, light-hole and bulk state populations based on Eqs.(9),(10) and (11).

## 10. Numerical validation of model

### 10.1. Verifying carrier sheet density conservation in the thermalization processes

The MB-MLME model includes a scheme for thermalizing carriers in both the well and bulk states to a common quasi-equilibrium Fermi-Dirac distribution. This is achieved using intra sub-band, inter sub-band as well as carrier leakage terms. In this section, we verify the accuracy of the MB-MLME thermalization model. We compare carrier distributions obtained from simulations of electrical pumping of a SQW active medium to analytic band filling calculations. Results from the two approaches agree well which verifies the accuracy of the MB-MLME model.

#### 10.1.1. Describing the Active Medium

The active medium used for all calculations consists of a single unstrained Gallium Arsenide (GaAs) Quantum Well with thickness $80A^0$ sandwiched between two $100A^0$ thick $Al_{0.2}Ga_{0.8}As$ barrier layers. For simplicity, parabolic sub-bands are assumed. The various quantized conduction electron ($E_{C(l)}^{qw-0}$), heavy-hole ($E_{HH(l)}^{qw-0}$) and light-hole ($E_{LH(l)}^{qw-0}$) energies are obtained by a solution of the 1-D Schrödinger's equation via the scattering matrix approach [36]. Our solutions yielded two conduction, heavy-hole and light-hole sub-bands each. The well depths, out of plane and in plane effective masses for the various conduction, heavy-hole and light-hole sub-bands are listed in Table.1.

| Material Parameter | Value |
| --- | --- |
| Well width ($L^{qw}$) | 95A$^0$ |
| Barrier width ($L^B$) | 100A$^0$ |
| Bandgap of well material ($E_G^{qw}$) | 1.42eV |
| Bandgap of barrier material ($E_G^B$) | 1.67eV |
| Out of-plane electron mass ($m_{C(l)\perp}^{qw}$) | 0.067m$_0$ (well) / 0.0836 m$_0$ (barrier) |
| Out of-plane heavy hole mass ($m_{HH(l)\perp}^{qw}$) | 0.377m$_0$ (well) / 0.39m$_0$ (barrier) |
| Out of-plane light hole mass ($m_{LH(l)\perp}^{qw}$) | 0.09m$_0$ (well) / 0.10m$_0$ (barrier) |
| In-plane conduction electron mass ($m_{C(l)\parallel}^{qw}$) | 0.067m$_0$ |
| In-plane heavy hole mass ($m_{HH(l)\parallel}^{qw}$) | 0.111m$_0$ |
| In plane light hole mass ($m_{LH(l)\parallel}^{qw}$) | 0.21m$_0$ |
| Unconfined electron mass in well ($m_C^{un}$) | 0.067 m$_0$ |
| Unconfined hole mass in well ($m_V^{un}$) | 0.34 m$_0$ |
| Quantized electron energy levels $E_{C(l)}^{qw-0}$ | 0.036eV, 0.10eV |
| Quantized heavy hole energy levels $E_{HH(l)}^{qw-0}$ | 0.0098eV, 0.0386eV |
| Quantized light hole energy levels $E_{LH(l)}^{qw-0}$ | 0.026eV, 0.09eV |

Table.1: Material parameters for GaAs/Al$_{0.2}$Ga$_{0.8}$As Quantum Well system

In the MB-MLME model, we span a broad transition energy range equivalent to 300nm starting from the energy bandgap $E_G^{qw}$ of the well material. This is to cover a broad optical bandwidth and encompass the entire energy bandwidth of excited carriers. This transition energy bandwidth was spanned by energy level pairs equally spaced in energy by an amount $\Delta E_t$=0.028eV. Such a value was chosen as it is comparable to the energy broadening due to dipole dephasing, suitable for generating smooth gain spectra. The above value of $\Delta E_t$ results in 18 pairs of energy levels spaced by $\Delta E^{qw}_{C-HH(l,i)} = \Delta E_t$, governing transitions between the $l=1$ conduction and $l=1$ heavy-hole sub-bands. Similarly, 18 pairs of levels governing transitions between the $l=1$ conduction and light-hole sub-bands are also spaced by $\Delta E^{qw}_{C-LH(l,i)} = \Delta E_t$. Thus, there are 18 levels in the first conduction, heavy-hole and light-hole sub-bands. The various transition energies ($E^{qw}_{C-HH(l,i)}$ etc.) as well as absolute energies ($E^{qw}_{C(l,i)}, E^{qw}_{HH(l,i)}$ etc.) for all levels are listed in Table.2. The second ($l=2$) conduction, heavy-hole and light-hole sub-bands are spanned by 15 levels each, with the same spacing between consecutive energy level pairs. For the bulk states, only the unconfined well states and barrier states are considered. It is also possible to include SCH states but we have not done so for simplicity. The bulk states are represented by a combined effective mass based on Eq.(7) using the effective mass values used in Table.1. The bulk states are spanned by 11 pairs of levels (11 each for conduction and valence band) which have the same spacing between consecutive energy level pairs $\Delta E^B_j = \Delta E_t$. Based on the energy level definitions and effective mass parameters in Table.1, dipole number densities are calculated according to Eq.3 and are listed in Table.2 along with all other MB-MLME model parameters such as transition times, dephasing rates and matrix elements.

For our simulations, we pump carriers into the bulk states via the $R_{pump}$ term in Eq.(17). The carrier rate equations are iterated for 10ns till steady state has been reached.

| MB-MLME model parameter | Value |
|---|---|
| Conduction-heavy hole transition energy $E^{qw}_{C-HH(l,i)}$ | [1.479, 1.507, 1.536, 1.564, 1.593, 1.621, 1.65, 1.678, 1.706, 1.735, 1.763, 1.792, 1.82, 1.848, 1.87, 1.905, 1.934, 1.962]eV *(l=1)* |
| | [1.572,1.6,1.628,1.657,1.685,1.714, 1.742,1.77,1.799,1.827,1.856,1.884, 1.912,1.941,1.969]eV (*l=2*) |
| Conduction-light hole transition energy $E^{qw}_{C-LH(l,i)}$ | [1.504,1.55,1.596,1.642,1.689 1.735,1.781,1.827,1.873,1.919 1.965,2.011,1.504,1.550,1.596 1.642,1.689,1.735,1.781,1.827 1.873,1.919,1.965 ,2.011] eV (*l=1*) |
| | [1.632,1.678,1.724,1.77,1.817 1.863,1.909,1.955,2.001,2.047 2.093,2.139,2.186,2.232,2.278]eV (*l=2*) |
| Bulk transition energy $E^B_j$ | [1.687,1.715,1.744,1.772,1.801 1.829,1.857,1.886,1.914,1.943 1.971]eV |
| Quantum well dipole number density $N^{qw-0}_{C(l,i)} = N^{qw-0}_{HH(l,i)} = N^{qw-0}_{LH(l,i)}$ | $4.95\times10^{15}$ m$^{-2}$ |
| Bulk dipole number density | [0.789 1.44 1.87 2.21 2.51 2.77 3.02 3.24 3.45 3.65 3.83]$\times10^{16}$ m$^{-2}$ |
| Intra sub-band EDT times | 50fs |
| Intra conduction sub-band EUT times | 99.11fs |
| Intra heavy hole sub-band EUT times | 75.56fs |
| Intra light hole sub-band EUT times | 76.93fs |
| Inter sub-band EDT times | 50fs |
| Inter conduction sub-band EUT times | 0.592ps |
| Inter heavy hole sub-band EUT times | 0.15ps |
| Inter light hole sub-band EUT times | 0.592ps |
| Heavy hole – light hole sub-band EUT times | 94.26fs (HH1-LH1), 1.11ps (HH1-LH2), 79.87fs (HH2-LH1), 0.37ps (HH2-LH2) |
| Carrier leakage EDT times | 0.5ps |
| Conduction band carrier leakage EUT time | 31.4ps (C1-B) |
| | 2.65ps (C2-B) |
| Valence band carrier leakage EUT time | 15.3ps(HH1-B) |
| | 5.098ps(HH2-B) |
| SQW Dipole dephasing rates | $3\times10^{13}$/s |
| Background refractive index ($n_r$) | 3.6 |
| Dipole matrix element for C-HH transitions ($|\mu_{C-HH^{qw}_{(l,i)\parallel}}|^2$)[20] | $\dfrac{e^2}{8\left(\omega^{qw}_{C-HH(l,i)}\right)^2}[1+\dfrac{E_{C(l)-HH(l)1}}{E^{qw}_{C-HH(l,i)}}]\cdot(28.8e)$ |
| Dipole matrix element for C-LH transitions ($|\mu_{C-LH^{qw}_{(l,i)\parallel}}|^2$) [20] | $\dfrac{e^2}{6\left(\omega^{qw}_{C-LH(l,i)}\right)^2}[\dfrac{5}{4}-\dfrac{3}{4}\dfrac{E_{C(l)-LH(l)}}{E^{qw}_{C-LH(l,i)}}]\cdot(28.8e)$ |
| Spontaneous decay lifetime for conduction-heavy hole transitions | $\left(n_r\left(\omega^{qw}_{C-HH(l,i)}\right)^3\left|\mu^{qw}_{C-HH(l,i)\square}\right|^2\right)^{-1}\pi\hbar\varepsilon_0 c^3$ |
| Spontaneous decay lifetime for conduction-light hole transitions | $\left(n_r\left(\omega^{qw}_{C-LH(l,i)}\right)^3\left|\mu^{qw}_{C-LH(l,i)\square}\right|^2\right)^{-1}\pi\hbar\varepsilon_0 c^3$ |
| Spontaneous decay lifetime for bulk transitions | 1ns |

Table.2. Calculated MB-MLME model parameters based on material parameters in Table.1

## *10.1.2. Analytical calculation of quasi-equilibrium carrier distributions*

The quasi-equilibrium Fermi-Dirac distributions can be calculated analytically based on the conservation of the total sheet carrier density $N_{2D}$ as shown in Eq.(30). A similar calculation was used in [24].

$$N_{2D}(\mu_C) = \left(n^{qw}+1\right)L^B\int_0^\infty \frac{\sqrt{2}\left(m_C^{barr}\right)^{3/2}}{\pi^2\hbar^3}\sqrt{\varepsilon}f_C(\varepsilon,\mu_C,T_L)d\varepsilon + n^{qw}L^{qw}\int_0^\infty \frac{\sqrt{2}\left(m_C^{un}\right)^{3/2}}{\pi^2\hbar^3}\sqrt{\varepsilon}f_C(\varepsilon,\mu_C,T_L)d\varepsilon$$
$$+n^{qw}\sum_{\forall l}\int_0^\infty \frac{\left(m^{qw}_{C(l)\square}\right)}{\pi\hbar^2}f_C(\varepsilon,\mu_C,T_L)d\varepsilon$$

(30)

In Eq.(30), the first term corresponds to the total carrier sheet carrier density due to all electrons in the conduction band of the bulk states. The second term corresponds to all electrons in the unconfined well bulk states. The third term corresponds to all electrons in all the sub-bands of the SQW. Note that all the integrands contain a common quasi-equilibrium Fermi-Dirac distribution $f_C(\varepsilon, \mu_C, T_L)$. In order to make comparisons to simulation results from the MB-MLME model, we first evaluate the total steady state sheet carrier density obtained by summing up the electron densities over all energy levels in the bulk and SQW states, i.e. $N_{2D} = \sum_j N_{C(j)}^B + \sum_l \sum_i N_{C(l,i)}^{qw}$. Then, we plug in this value of $N_{2D}$ into the LHS of Eq. (30) to obtain the analytical carrier distribution. In Fig.7 (a)-(d), we plot the steady state carrier occupational distributions obtained from MB-MLME model simulations as well as the corresponding analytic calculations.

It can be seen in Fig.7 that the carrier occupational distributions obtained from the MB-MLME model for the bulk states (Fig.7a), second conduction sub-band-C2 (Fig.7b) and first conduction sub-band-C1 (Fig.7c) have thermalized to a common Fermi-Dirac distribution as expected from the formulations in Section 7. Furthermore, the analytic calculations in Fig.7d based on Eq.(30) matches very well with the simulated curves. This verifies that the various intra sub-band, inter sub-band and carrier leakage terms act to thermalize excited electrons to the correct Fermi-Dirac distribution. The case for holes follows similarly.

*10.2. Quantum Well Gain Spectra Simulations*

In this section, we use a 2-D implementation of the MB-MLME model as well as FDTD from section 9 to simulate SQW gain spectra for the system described in Section 10.1. We make comparisons to bulk GaAs gain spectra obtained from the MLME model [17] to distinguish the gain dispersions. The value of the SQW absorption spectrum at zero excitation shows good quantitative agreement with the results from [37] alluding to the numerical accuracy of the MB-MLME approach. All the MB-MLME model parameters are the same as in Tables. 1 and 2. For simulating bulk GaAs gain spectra, we use the same parameters as in [17] which have shown to agree well with previous work [37]. In order to obtain the gain spectra, we launch a short pulse into a 0.3µm wide waveguide uniformly filled with active medium and compare the pulse spectra between the output and input points. A FDTD spatial resolution of $\Delta x = \Delta z = 20 nm$ and temporal resolution $\Delta t = .023 fs$ which satisfies the Courant stability condition was used for the simulations. The MB-MLME carrier rate equations were evaluated once every $\kappa = 10$ iterations to improve computational efficiency. The results are plotted in Fig.8.

In Fig.8a, the SQW gain spectra at various levels of excitation are plotted. The sheet carrier densities in curve(1) >curve(2)>curve (3). The step like dispersion due to the step like density of states can easily be discerned in the gain curve at zero excitation (curve (3)). On the contrary, the bulk gain spectrum from the MLME model at zero excitation in Fig.8b does not exhibit such step-like dispersion but instead has a dispersion proportional to $\sqrt{\varepsilon}$, in line with the bulk semiconductor density of states. Thus, only the MB-MLME model with a separate treatment of various sub-bands can account for these SQW characteristics. The magnitude of the gain/absorption coefficient at zero excitation in Fig.8a matches well with results from [37], indicating that the various effective masses, dipole matrix elements and dephasing rates used are sastisfactory. The lorentzian broadening used in the MB-MLME model results in some spurious absorption below the band-edge as seen in Figs.8a and 8b. A more accurate treatment of dipole dephasing, using non Markovian relaxation [38] is thus necessary. Note in Fig.8a that with increasing excitation, the degree of absorption reduces and eventually gain is achieved. At lower levels of excitation, only the first sub-bands are sufficiently filled and gain is only obtained at these energies (curve (2), Fig.8a).

However, at higher levels of excitation, the higher sub-bands are sufficiently filled as evident in the appearance of 'steps' in the gain spectrum (curve (3), Fig.8a). The peak gain increases and shows blue shift w.r.t energy. This is the well-known band filling effect associated with semiconductor media [37]. In Fig.8b, while the peak gain increases and shows a blue shift with increasing excitation, similar to the case of the SQW gain spectra in Fig.8a, a step like dispersion is not observed due to the absence of higher sub-bands as in the SQW case.

### *10.3. Photonic Crystal Laser Simulations*

In this section, we show that the developed MB-MLME FDTD approach is computationally efficient enough to perform 2-D simulations of nanophotonic devices of complex structural geometry by applying it to nanobeam photonic crystal cavity lasers. The device schematic used for the simulations is presented in Fig.9a. It consists of 0.5µm wide, 6 µm long waveguide with air holes of diameter 350 nm periodically spaced by 450 nm. The periodicity is disturbed by the absence of a single hole at the center of the waveguide which forms the required cavity for lasing. Light is extracted at the right end of the waveguide in Fig.9a. The effective index of the waveguide is adjusted to 3.175 so as to obtain a cavity with resonant wavelength close to ~1540 nm. The surrounding cladding space consists of air with a total simulation area of 2µm×6µm. The waveguide (barring the holes) is filled uniformly with SQW active media.

The SQW system here consists of 6 95$A^0$ thick AlGaInAs compressively-strained wells and 7 100 $A^0$ thick tensile-strained AlGaInAs barrier layers. The thicknesses and compositions were chosen for peak gain around 1550nm. The material band-gaps including strain corrections were obtained from [39]. While an accurate calculation of the strained band structures requires the use of the *'k.p'* approach, here for the sake of simplicity we treat the strained structures under parabolic approximations. However, as mentioned previously, the MB-MLME model can negotiate more complicated band structures. The various quantized energy levels for conduction electrons, heavy-holes and light-holes are calculated similar to Section 10.1 using the scattering matrix approach [36]. There are one conduction and two heavy-hole sub-bands. The light-hole sub-bands are located at high enough energies so they may be neglected in our simulations. Bulk-barrier and unconfined-well states are considered via a single effective mass according to Eq.(7). We consider the carrier dynamics associated with the bulk states but not the dynamics of the bulk dipoles themselves as explained in Section 4. The various MB-MLME model active medium parameters are presented in Table.3.

| Material Parameter | Value |
| --- | --- |
| Well width ($L^{qw}$) | 95A$^0$ |
| Barrier width ($L^B$) | 100A$^0$ |
| Bandgap of well material ($E_G^{qw}$) | 0.76eV |
| Bandgap of barrier material ($E_G^B$) | 0.89eV |
| Out of-plane electron mass ($m_{C(l)\perp}^{qw}$) | 0.0462$m_0$ (well), 0.0589$m_0$ (barrier) |
| Out of-plane heavy hole mass ($m_{HH(l)\perp}^{qw}$) | 0.3192$m_0$ (well), 0.3601$m_0$ (barrier) |
| Out of-plane light hole mass ($m_{LH(l)\perp}^{qw}$) | 0.0586$m_0$ (well), 0.0765$m_0$ (barrier) |
| In-plane electron mass ($m_{C(l)\parallel}^{qw}$) | 0.0462$m_0$ |
| In-plane heavy hole mass ($m_{HH(l)\parallel}^{qw}$) | 0.072$m_0$ |
| Bulk electron mass ($m_C^{un}, m_C^{barr}$) | 0.0462$m_0$, 0.0559$m_0$ |
| Bulk hole mass ($m_V^{un}, m_V^{barr}$) | 0.3192$m_0$, 0.0723$m_0$ |
| Quantized electron energy ($E_{C(l)}^{qw-0}$) | 0.0267eV |
| Quantized heavy hole energy ($E_{HH(l)}^{qw-0}$) | 0.0098eV, 0.036eV |
| Number of conduction sub-band levels | 12($l$=1) |
| Number of heavy hole sub-band levels | 12($l$=1), 7($l$=2) |
| Number of bulk levels | 7 |
| Energy level separation $\Delta E_{C(l,i)}^{qw}$ | 0.01eV($l$=1) |
| Energy level pair separation $\Delta E_{HH(l,i)}^{qw}$ | 0.0064eV($l$=1) <br> 0.0082eV($l$=2) |
| Bulk energy level pair separation $\Delta E_j$ | .0165eV |
| Intra sub-band EDT times | 50fs |
| Inter sub-band EDT times | 50fs |
| Carrier leakage EDT times | 0.5ps |
| SQW Dipole Dephasing rates | $4\times10^{13}$/s |
| Dipole matrix element for C-HH transitions ($\|\mu_{C-HH(l,i)\|}^{qw}\|^2$)[20] | $\frac{e^2}{8\left(\omega_{C-HH(l,i)}^{qw}\right)^2}[1+\frac{E_{C(l)-HH(1)}}{E_{C-HH(l,i)}^{qw}}]\cdot(20e)$ |
| Dipole matrix element for C-LH transitions $\|\mu_{C-LH(l,i)\|}^{qw}\|^2$ [20] | $\frac{e^2}{6\left(\omega_{C-LH(l,i)}^{qw}\right)^2}[\frac{5}{4}-\frac{3}{4}\frac{E_{C(l)-LH(l)}}{E_{C-LH(l,i)}^{qw}}]\cdot(20e)$ |

Table. 3. Material and MB-MLME model parameters for AlGaInAs system used in photonic crystal nanobeam cavity laser simulations.

The FDTD simulations are performed at a spatial resolution of $\Delta x = \Delta z = 20 nm$ and a temporal resolution of $\Delta t = 0.046 fs$. The MB-MLME carrier rate equations were evaluated every $\kappa = 10$ iterations as discussed in Section 9 to improve computational efficiency. The active medium was pumped at various current density values $J$. A very weak seed pulse was impinged on the structure to initiate the lasing process. After executing the FDTD simulation for a total time of 100 ps or ~2 million iterations, steady-state was reached. At steady-state, lasing was seen with the defect serving as the lasing cavity as seen in the snapshot of the electric field $E_x$ Fig.9b.

*10.4. Notes on Computational Efficiency*

The computational overhead of the MB-MLME FDTD approach depends on the number of sub-bands as well as the number of levels per sub-band. Typical SQW systems would have about 2-3 hole (heavy plus light-hole) sub-bands, 1-2 conduction sub-bands and also bulk state sub-bands. A good measure of computational efficiency can be the overhead of active medium FDTD simulations over passive medium FDTD. Since the MLME model has already been shown to be computationally efficient and applied successfully to 2-D [16]-[17] and 3-D [18] simulations, a good measure of computational efficiency is the overhead of the MB-MLME model over the MLME model. An MB-MLME model with 6 SQW sub-bands and 2 bulk bands with 18 levels for each $l=1$, 15 for each $l=2$ sub-band and 11 levels each for bulk conduction and valence bands identical to section 10.1 and 10.2 was used. The corresponding MLME model for comparison contained 18 levels each for 1 conduction and 1 valence band. The resulting overhead in this case was ~1.5 times. If the total number of SQW sub-bands was reduced to 3 as in the system described in Section 10.3, the overhead compared to the corresponding to the MLME model was 1.2. Thus, the MB-MLME model is of similar computational efficiency to the MLME model. In addition, using different time steps for the active medium variable update can further reduce simulation times. For the general 3-D case, in addition to the 6 variables for electromagnetic fields, additional memory requirements include a maximum of $2.(l+2).i$ variables for carrier density parameters for $l$ SQW sub-bands with $i$ levels each and $(l+2).i$ variables for each polarization density vector component. Thus the maximum memory overhead over passive FDTD would be $\frac{3(l+2)i+6}{6}$; For $l=3$ and $i=20$, this is ~50 times. While, the overhead appears large, it would still be smaller than the case of using finely resolved momentum states and dynamically changing chemical potentials.

## 11. Conclusion

In this paper, an active medium model to govern the light-matter interaction with SQW was formulated. This model was called the MB-MLME model. It was physically accurate and also computationally efficient enough to be applied to FDTD simulations of Nanophotonic devices of complex 2-D/3-D geometry. The concept of representing several transverse momentum states within each SQW sub-band with few multi-electron states, resulted in computational efficiency by circumventing the need for tracking carrier and dipole dynamics over finely resolved transverse momentum states.

Additionally, a scheme of intra sub-band, inter sub-band and carrier leakage terms based on EDT and EUT was developed which automatically thermalized the carrier distribution to a common quasi-equilibrium Fermi-Dirac distribution. These terms avoided numerically cumbersome iterative procedures. This further augmented the computational efficiency and enabled FDTD simulations of complex device geometries. In order to verify the developed MB-MLME model, we compared quasi-equilibrium distributions obtained from electrically pumping a SQW system to analytical band filling calculations. Good agreement was demonstrated, thus verifying the accuracy of the scheme of transitions presented. Additionally, SQW gain spectra were evaluated using the MB-MLME FDTD model which showed characteristic SQW gain spectra features. The absorption spectra also agreed quantitatively with published literature.

The utility of the MB-MLME FDTD approach in simulating complex photonic structures was demonstrated by applying it to a nanobeam photonic crystal laser. It was seen that the MB-MLME takes only 1.2-1.5 times compared to the MLME model. Since the MLME model has already been successfully applied to complex 2-D and 3-D simulations with an overhead of 2-3x [16]-[17], it follows that the MB-MLME model is equally applicable to such structures. Future work could involve the investigation of lasing dynamics in complex geometries such as nanoplasmonic, photonic crystal cavities etc.

**Appendix. A**

In this section, we describe the assignment of various energy levels in the MB-MLME model. First we define the maximum number of allowed levels per sub-band $i_{max}$ and the total optical energy bandwidth of interest $\Delta E_{optical}$. $\Delta E_{optical}$ must be chosen so that the carrier occupational probabilities converge to zero at the highest energies. Then, the spacing between consecutive level-pairs is $\Delta E_t = \Delta E_{optical}/i_{max}$. Next, the various quantized energy $E_{C(l)}^{qw-0}$, $E_{HH(l)}^{qw-0}$ or sub-band edges are calculated using the Schrödinger's equation. Since, only $\Delta l=0$ transitions are considered, the conduction to heavy-hole sub-band edge transition energies are given by $E_{C(l)-HH(l)} = E_{C(l)}^{qw-0} + \left|E_{HH(l)}^{qw-0}\right| + E_G^{qw}$. Since interaction of light with each sub-band approximately stops at this energy, we define the lower limit of the energy bracket spanned by the first transition level-pair in the $l^{th}$ sub-bands to be $E_{C-HH(l,1)}^{qw-} = E_{C(l)-HH(l)}$. Since various energy level-pairs are spaced $\Delta E_t$ apart, in general we have Eq. (A.1).

$$E_{C-HH(l,i)}^{qw-} = E_{C-HH(l,i)}^{qw-} + \Delta E_t \cdot (i-1) \tag{A.1}$$

The corresponding upper limit of the energy bracket spanned by the $i^{th}$ transition level pair in the $l^{th}$ sub-band would then be $E_{C-HH(l,i)}^{qw+} = E_{C-HH(l,i)}^{qw-} + \Delta E_t$. Since, the transition energy between conduction and heavy-hole sub-bands $E_{C-HH_{(l,i)}^{qw}}$ is centered in the energy bracket $[E_{C-HH(l,i)}^{qw-}, E_{C-HH(l,i)}^{qw+}]$, it is given by $E_{C-HH(l,i)} = (E_{C-HH(l,i)}^{qw-} + E_{C-HH(l,i)}^{qw+})/2$. The corresponding absolute energies of the various conduction sub-band multi-electron states are obtained by conserving transverse momentum according to Eq. (A.2).

$$E_{C(l,i)}^{qw}(k_t) + E_{HH(l,i)}^{qw}(k_t) = E_{C-HH(l,i)}^{qw} \tag{A.2}$$

Each conduction sub-band also interacts with a light-hole sub-band, therefore the conduction to light-hole transition energies are dependent on the conduction to heavy-hole transition energies. We obtain the definitions of $E^{qw}_{C-LH(l,i)}$ by simultaneously solving Eqs. (A.2) and (A.3).

$$E^{qw}_{C(l,i)}(k_t) + E^{qw}_{LH(l,i)}(k_t) = E^{qw}_{C-LH(l,i)} \tag{A.3}$$

It is possible that that there are more number of heavy/light-hole sub-bands than conduction sub-bands. In such cases, their absolute energies can be spaced by $\Delta E_t / 2$, i.e.

$$E^{qw}_{HH/LH(l,i)} = E^{qw-0}_{HH/LH(l)} + (i-1)\Delta E_t / 2.$$

**Appendix.B**

When the carrier occupational probability distribution has been thermalized, it implies that net intraband transitions between any two of upper ($u$) and lower levels ($d$) have to approach 0. We mathematically express this in Eq.(B.1). Note that we do not place any constraints on the EDT and EUT times $\tau_d$ and $\tau_u$ respectively. They may be functions of carrier number density, carrier temperature, lattice temperature etc.

$$\frac{N_u(\mathbf{r},t)\left(1 - \frac{N_d(\mathbf{r},t)}{N_d^0}\right)}{\tau_d(\mathbf{r},t)} - \frac{N_d(\mathbf{r},t)\left(1 - \frac{N_u(\mathbf{r},t)}{N_u^0}\right)}{\tau_u(\mathbf{r},t)} = 0 \tag{B.1}$$

In a quasi-equilibrium situation, in addition to the condition in Eq.(B.1), the distribution must also follow a Fermi-Dirac function (since electrons are fermions) and hence we have the additional requirement of Eq.(B.2).

$$\left.\frac{N_{u(d)}(\mathbf{r},t)}{N_{u(d)}^0}\right|_{quasi-equilibrium} = \left(1 + \mathbf{exp}\left\{\left(E_{u(d)} - \mu(\mathbf{r},t)\right)/k_B T_e(\mathbf{r},t)\right\}\right)^{-1} \tag{B.2}$$

If we substitute Eq.(B.2) in Eq. (B.1), we obtain the general expression for the ratio between EDT and EUT times for any two levels in Eq.(B.3) while eliminating the need to calculate the chemical potential $\mu(\mathbf{r},t)$.

$$\tau_u(\mathbf{r},t) = \frac{N_d^0}{N_u^0} \tau_d(\mathbf{r},t) \exp\left\{\frac{E_u - E_d}{k_B T_e(\mathbf{r},t)}\right\} \tag{B.3}$$

**Appendix.C**

| Parameter | Definition |
|---|---|
| $\alpha = \{HH, LH\}$ | |
| $E_{C/\alpha(l)}^{qw-0}$ | Quantized energy levels of electrons in the conduction /$\alpha$ ($l^{th}$ sub-band edge energy) |
| $E_{C-\alpha(l,i)}^{qw}$ | Transition energy for the $i^{th}$ conduction – $\alpha$ level pair in the $l^{th}$ SQW sub-band |
| $\omega_{C-\alpha(l,i)}^{qw}$ | Angular frequency corresponding to transition energy $E_{C-\alpha(l,i)}^{qw}$ |
| $\lambda_{C-\alpha(l,i)}^{qw}$ | Wavelength corresponding to transition energy $E_{C-\alpha(l,i)}^{qw}$ |
| $\Delta E_{C-\alpha(l,i)}^{qw}$ | Transition energy separation between consecutive level pairs. |
| $E_{C-\alpha(l,i)}^{qw\pm}$ | Upper (lower) limit of conduction-$\alpha$ transition energy bracket of quantum wells. |
| $E_{C/\alpha(l,i)}^{qw}$ | Absolute energy of the $i^{th}$ conduction/$\alpha$ energy level in the $l^{th}$ SQW sub-band (w.r.t vacuum) |
| $\Delta E_{C/\alpha(l,i)}^{qw}$ | Energy separation between consecutive conduction/$\alpha$ sub-band energy levels of SQW. |
| $E_{C/\alpha(l,i)}^{qw\pm}$ | Upper(lower) limit of conduction/$\alpha$ sub-band multi electron state in SQW. |
| $\vec{P}_{C-\alpha(l,i)}^{qw}(\mathbf{r},t)$ | Polarization density vector for SQW corresponding to transition energy $E_{C-\alpha(l,i)}^{qw}$ |
| $N_{C/\alpha(l,i)}^{qw}(\mathbf{r},t)$ | Number of carriers per unit area in the $i^{th}$ energy level of the $l^{th}$ conduction/$\alpha$ SQW sub-band |
| $N_{C/\alpha(l,i)}^{qw-0}(\mathbf{r})$ | Total number of states per unit area in the $i^{th}$ level of the $l^{th}$ conduction/$\alpha$ SQW sub-band |

| Symbol | Description |
|---|---|
| $N^{qw}_{dip(l,i)}$ | SQW dipole number density |
| $m^{qw}_{C/\alpha(l)\parallel}$ | Effective in-plane mass of $l^{th}$ conduction/$\alpha$ sub-band of SQW |
| $m^{qw}_{C/\alpha(l)\perp}$ | Effective out of plane mass of $l^{th}$ conduction/$\alpha$ sub-band of SQW |
| $\|\mu^{qw}_{C-\alpha(l,i)\parallel}\|^2$ | In-plane Dipole Matrix Element of SQW |
| $\|\mu^{qw}_{C-\alpha(l,i)\perp}\|^2$ | Out-of plane Dipole Matrix Element of SQW |
| $\tau^{qw}_{C-\alpha(l,i)}$ | Spontaneous recombination lifetimes between level pairs with transition energy $E^{qw}_{C-\alpha(l,i)}$ of SQW |
| $\tau_{C/\alpha(l,[i,j])}$ | Intra sub-band transition times from level $i$ to $j$ in the $l^{th}$ conduction/$\alpha$ SQW sub-band |
| $\tau_{C/\alpha([l,m],1)}$ | Inter sub-band transition times from SQW conduction/$\alpha$ sub-band $l$ to $m$ |
| $\tau_{C/\alpha(l)\rightarrow B}$ | Electron(heavy, light-hole) escape time to bulk states |
| $\tau_{B\rightarrow C/\alpha(l,1)}$ | Electron(heavy, light-hole) capture time to SQW states |
| $E^B_j$ | Transition energy for the $j^{th}$ pair of bulk levels |
| $\Delta E^B_j$ | Transition energy separation between consecutive bulk level pairs. |
| $\omega^B_j$ | Angular frequency corresponding to $E^B_j$ |
| $\lambda^B_j$ | Wavelength corresponding to $E^B_j$ |
| $E^{B\pm}_j$ | Upper(lower) transition energy limit of energy bracket centered $E^B_j$ |
| $E^B_{C/V(j)}$ | Absolute energy of the $j^{th}$ conduction(valence) bulk energy level |
| $\Delta E^B_{C/V(j)}$ | Energy separation between consecutive conduction(valence) bulk energy levels |
| $E^{B\pm}_{C/V}$ | Upper(lower) energy limit of the $j^{th}$ conduction(valence) multi-electron bulk state. |
| $\vec{P}^B_j(\mathbf{r},t)$ | Polarization density vector for bulk corresponding to transition energy $E^B_j$ |
| $\tau_{(i,j)C/V}$ | Intraband transition times between the $i^{th}$ and $j^{th}$ levels of the conduction (valence) band. |
| $\varepsilon_b$ | Background permittivity |
| $\Gamma$ | Optical confinement factor |

# Figures and Captions

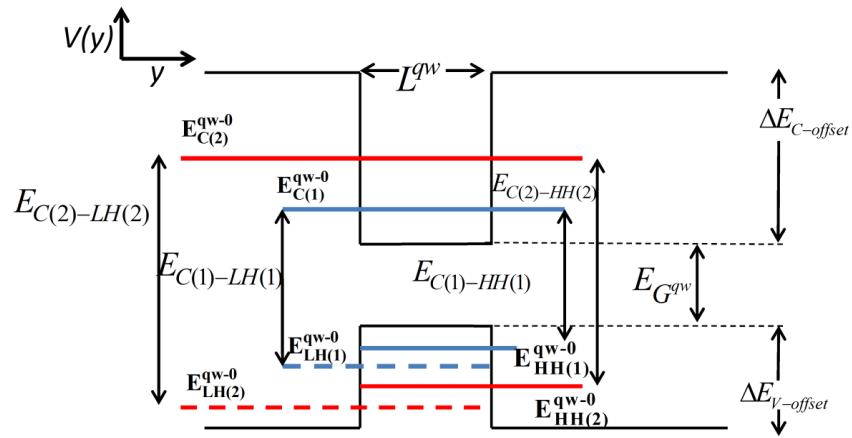

Fig.1. shows the quantization of energy states in a SQW. The square well potential results in the quantization of the energy levels that a conduction band electron can occupy. The first quantized conduction band energy levels are represented by blue lines and second quantized states represented by red lines. Similarly, heavy-hole (solid) and light-hole (dashed) states are also quantized. See Fig.2 for the corresponding energy-momentum dispersions.

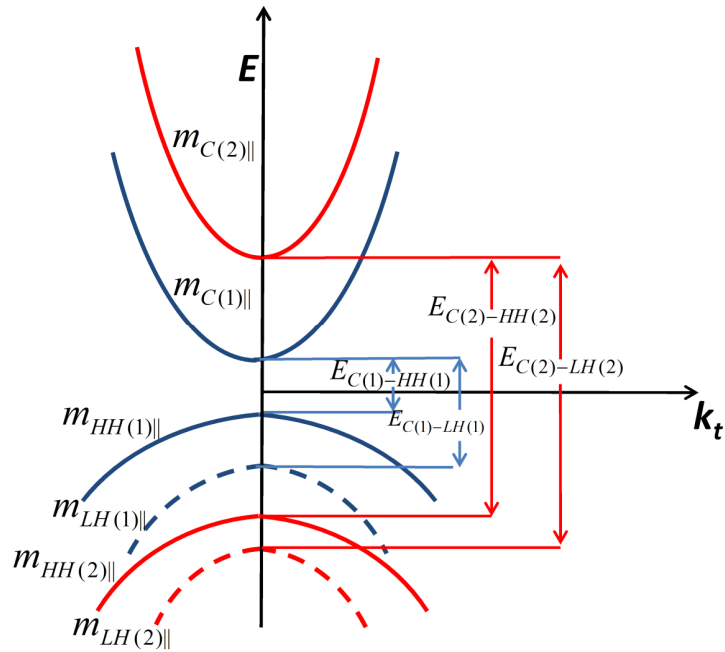

Fig.2: Energy –momentum dispersions for electrons in SQW are a series of parabolic sub-bands which represent the fact that electrons are quantized due to the square potential in the y direction but are 'free' to move in the plane of the wells in the *x* and *z* directions.

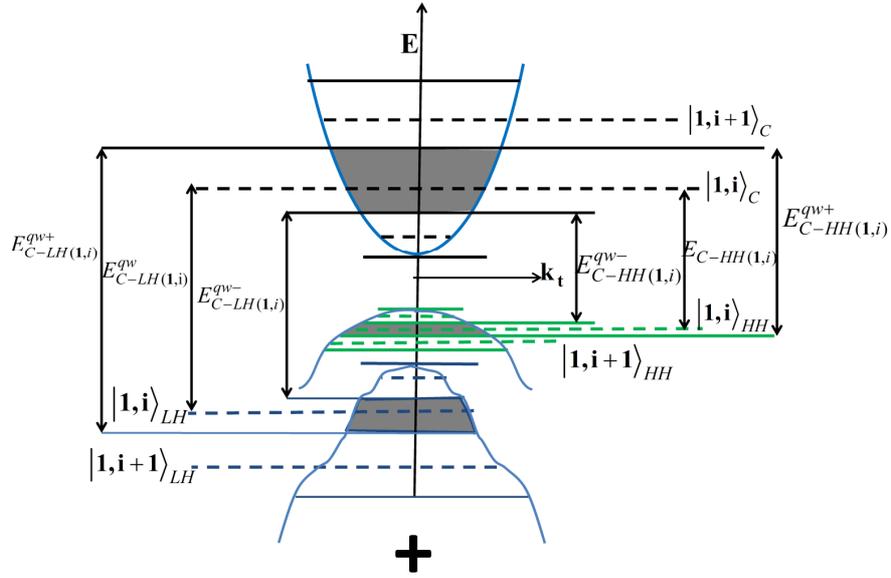

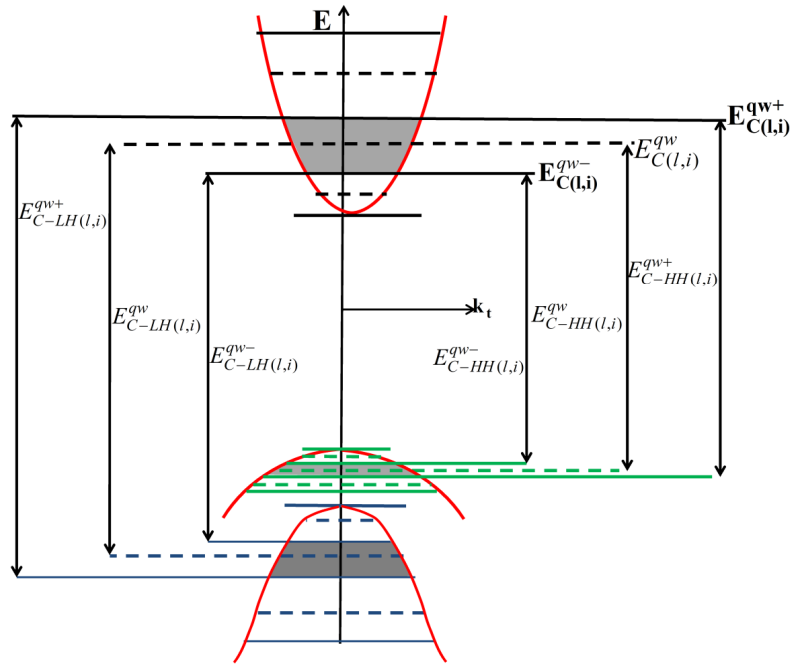

Fig.3. In the MB-MLME model, several momentum states within each sub-band $l$ are represented by a single broadened multi-electron state as shown by the shaded bands above. The overall system is then an ensemble of the various $l$ sub-bands, each spanned by a series of broadened multi-electron states. Since the number of sub-bands are few in number whereas the number of momentum states are large in number, significant computational savings can be achieved by this approach. Here, the various conduction and valence sub-bands with the same $l$ are depicted grouped together as transitons in SQW can be approximated (although not necessary) to follow the $\varDelta l=0$ constraint.

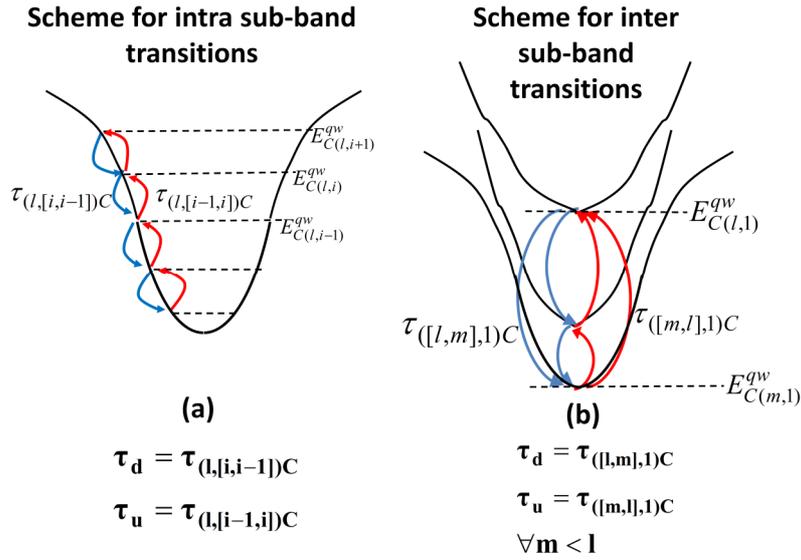

**Scheme for intra sub-band transitions**

(a)

$\tau_d = \tau_{(l,[i,i-1])C}$

$\tau_u = \tau_{(l,[i-1,i])C}$

**Scheme for inter sub-band transitions**

(b)

$\tau_d = \tau_{([l,m],1)C}$

$\tau_u = \tau_{([m,l],1)C}$

$\forall m < l$

Fig.4a depicts the scheme of intra sub-band transitions. A series of energy up transitions (EUT) (red) with transition time $\tau_u$ and energy down transitions (EDT) (blue) with transition times $\tau_d$ between two levels within the sub-band can be used to automatically thermalize carriers to the correct Fermi-Dirac distributions without the need for iterative chemical potential calculations. Fig.4b depicts the scheme of inter sub-band transitions. EUT (red) and EDT (blue) between the bottom most levels of two sub-bands can be used to thermalize carriers between two sub-bands to a common Fermi-Dirac distribution. Thus, when all sub-bands are coupled by such EDT and EUT, they will thermalize to a common quasi-equilibrium Fermi-Dirac distribution.

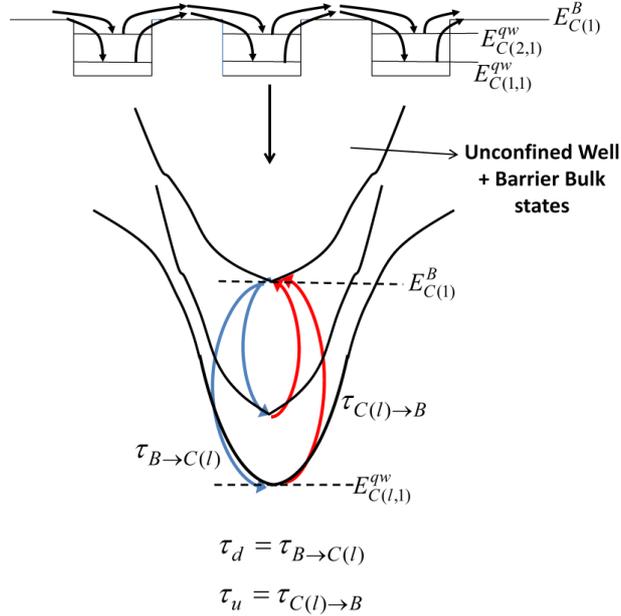

$\tau_d = \tau_{B \to C(l)}$

$\tau_u = \tau_{C(l) \to B}$

Fig.5 shows how carrier leakage to bulk states and the process of carrier capture back into the wells is included via a scheme of EUT and EDT between bulk and SQW states. This leads to SQW and bulk populations to be thermalized to a common Fermi-Dirac distribution.

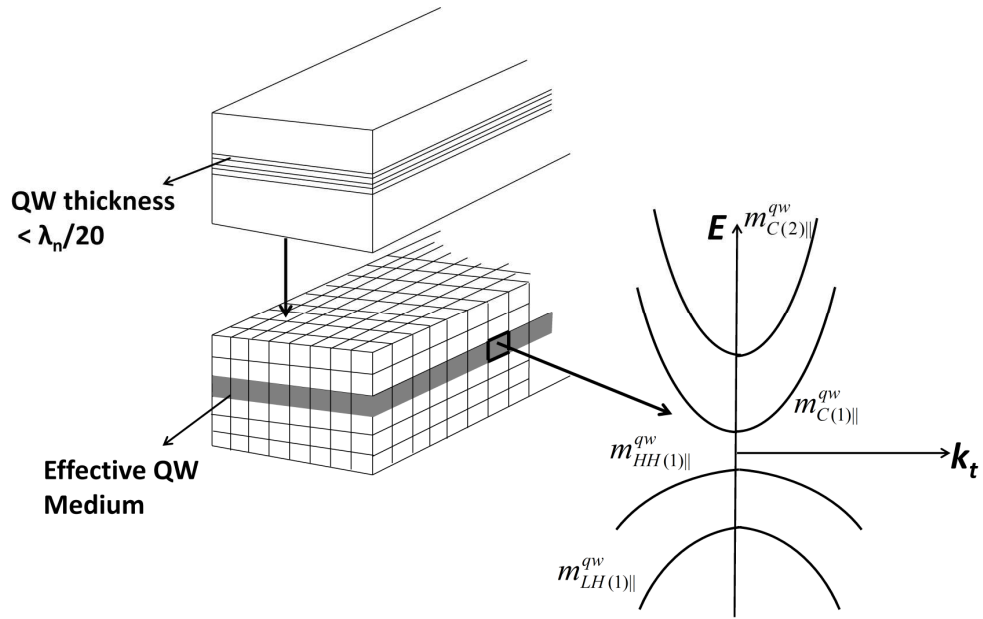

Fig.6. Since the SQW width is much smaller than the typical FDTD spatial grid size, we represent the entire SQW active medium by a uniform effective medium with an effective bandstructure containing details of the barrier as well as well states.

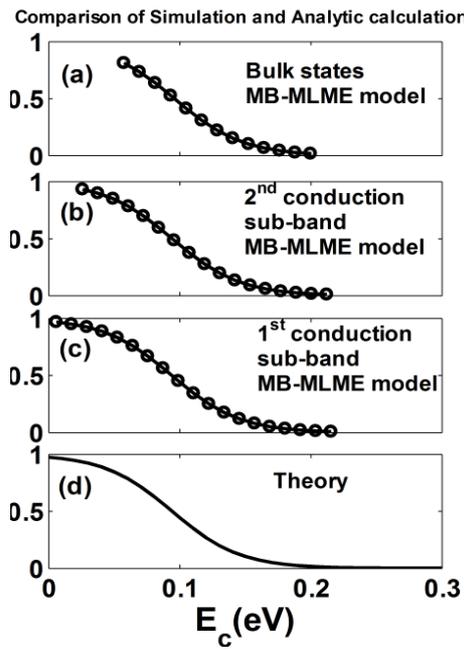

Fig.7 (a)-(d) show plots of quasi-equilibrium carrier occupational distributions obtained from the MB-MLME model for (a) the bulk states (b) 2nd conduction sub-band and (c) first conduction sub-band state . The carrier occupational distributions in each sub-band as well as the bulk states converge to a common Fermi-Dirac distribution which agrees with the analytic calculations (d). The MB-MLME calculations concur with the analytical results, verifying the scheme of the proposed thermalizing terms.

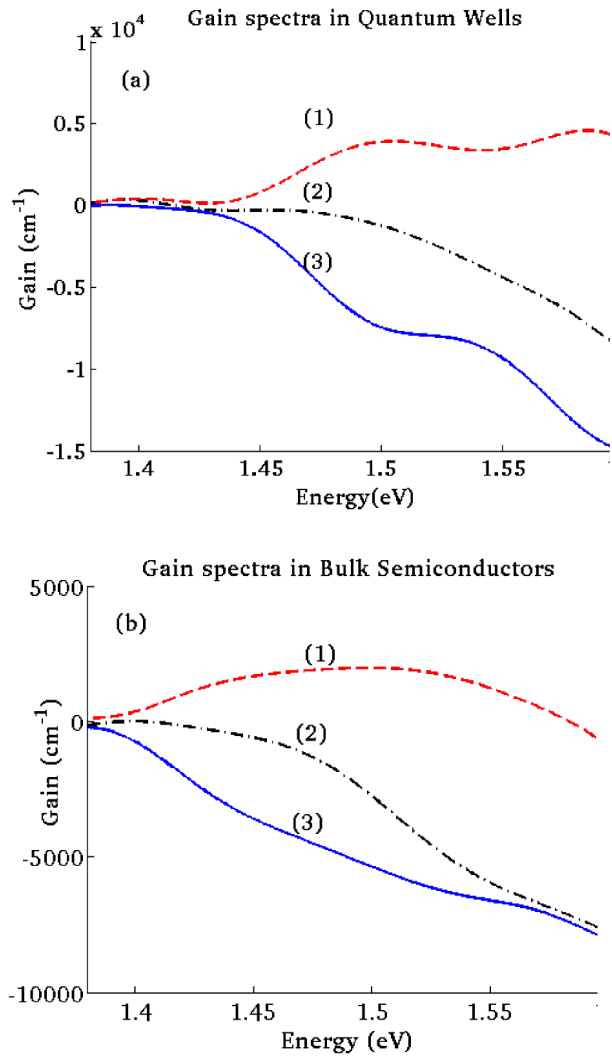

Fig.8a shows simulated gain spectra for GaAs based quantum well active media by passing a pulse through a waveguide. The step like shape corresponding to the step like density of states is clearly seen in curve(3) which is completely in the ground state. The carrier density in curve (1)> curve (2)>curve (3). Fig.8b shows gain spectra for bulk GaAs with a shape corresponding to the bulk density of states. The gain spectra without any excitation, i.e curve(3) shows good quantitative agreement with results from [37].

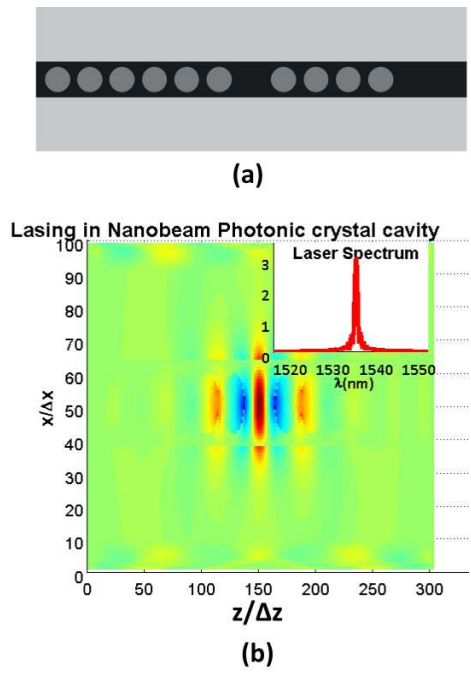

Fig.9 (a). A 2-D device schematic for nanobeam photonic crystal laser. The device consists of a 500nm waveguide with 350nm diameter air holes along the waveguide's length with period 450nm. The removal of a single period serves as a defect to form the lasing cavity.(b)Snapshot of steady state electric field component $E_x$ showing steady state lasing with defect serving as cavity.